\documentclass[12pt, leqno]{article}
 \usepackage[toc,page]{appendix}
\usepackage{amsmath,setspace,multirow,lineno}
\usepackage[font=small,labelfont=bf]{caption}
\usepackage[top=1.2in, bottom=1.2in, left=1.2in, right=1.2in]{geometry}

\usepackage[round]{natbib}
\usepackage{color,soul}

\DeclareCaptionStyle{italic}[justification=centering]{labelfont={bf},textfont={it},labelsep=colon}
\usepackage{graphicx,psfrag,epsf,textcomp,epstopdf, amsthm}
\usepackage{enumerate, dsfont}
\usepackage{natbib}
\usepackage{url,xcolor}
\usepackage{booktabs, subfig, bm, paralist,mathpazo,tikz,todonotes,longtable,microtype}
\usepackage[linesnumbered,ruled,vlined]{algorithm2e}

\usepackage[pdftex,colorlinks=true]{hyperref}
\definecolor{darkblue}{rgb}{0,0,.6}
\hypersetup{citecolor=darkblue,linkcolor=darkblue,urlcolor=darkblue}
\definecolor{DarkRed}{rgb}{.7,0,.4}

\usepackage{comment,orcidlink}

\newcommand{\blind}{0}

\newcommand{\X}{\mathcal{X}}

\graphicspath{{plots/}}

\newcommand{\Rlogo}{\protect\includegraphics[height=1.8ex,keepaspectratio]{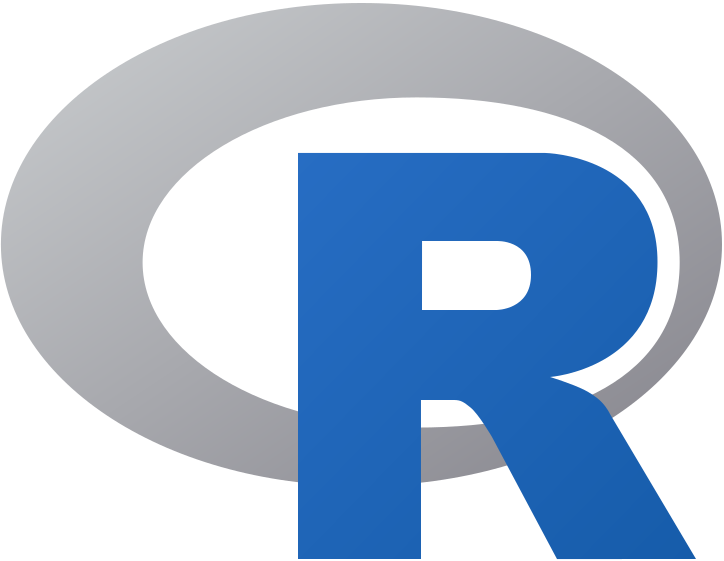}}

\newsavebox\CBox

 \newtheorem{@definition}{\sc Definition}[section]

  \renewcommand\X{\mathcal{X}}

\date{}

\begin{document}

\def\spacingset#1{\renewcommand{\baselinestretch}{#1}\small\normalsize} \spacingset{1}

\if0\blind
{
\title{\bf A robust scalar-on-function logistic regression for classification}}
\author{
Muge Mutis \orcidlink{0000-0002-9801-4835} \footnote{Corresponding address: Graduate School of Natural and Applied Sciences, Yildiz Technical University, 34220 Esenler-Istanbul, Turkey; Email: mugemutis@gmail.com} \\
Graduate School of Natural and Applied Sciences \\
Yildiz Technical University \\
\\
Ufuk Beyaztas \orcidlink{0000-0002-5208-4950} \\
Department of Statistics \\
Marmara University \\
\\
Gulhayat Golbasi Simsek \orcidlink{0000-0002-8790-295X} \\
    Department of Statistics \\
    Yildiz Technical University \\
    \\
Han Lin Shang \orcidlink{0000-0003-1769-6430} \\
    Department of Actuarial Studies and Business Analytics \\
    Macquarie University 
}
\maketitle
\fi

\if1\blind
{
\title{\bf A robust scalar-on-function logistic regression for classification}
} \fi

\maketitle

\begin{abstract}
Scalar-on-function logistic regression, where the response is a binary outcome and the predictor consists of random curves, has become a general framework to explore a linear relationship between the binary outcome and functional predictor. Most of the methods used to estimate this model are based on the least-squares type estimators. However, the least-squares estimator is seriously hindered by outliers, leading to biased parameter estimates and an increased probability of misclassification. This paper proposes a robust partial least squares method to estimate the regression coefficient function in the scalar-on-function logistic regression. The regression coefficient function represented by functional partial least squares decomposition is estimated by a weighted likelihood method, which downweighs the effect of outliers in the response and predictor. The estimation and classification performance of the proposed method is evaluated via a series of Monte Carlo experiments and a strawberry puree data set. The results obtained from the proposed method are compared favorably with existing methods.

\vspace{.1in}\noindent \textit{Keywords}: Basis function expansion; Functional partial least squares; Robust estimation; Strawberry purees; Weighted likelihood.
\end{abstract}

\newpage
\spacingset{1.58} 

\section{Introduction} \label{sec:intro}

With the development of technology in data collection tools and the widespread availability of functional data, functional data analytic tools have been widely used in many branches of science with a range of applications. Among many others, the scalar-on-function logistic regression (FLogR) proposed by \cite{Ratcliffe}, in which the response is a binary outcome, and predictor consists of random curves, has become a general framework to investigate the relationship between the predictor and the response and to classify the binary outcome.

Several methods, such as generalized functional linear model \citep{James2002, Muller2005, Goldsmith2011, Ogden2010}, functional principal component (FPC) regression \citep{Ratcliffe, Escabias2005, LengMuller, Aguilera2008, WeiTang}, penalized $\log$-likelihood \citep{Goldsmith2011}, and LASSO \citep{MousaviSorensen} have been proposed to estimate the regression coefficient function in the FLogR model. Recently, \cite{Mousavi2018} compared the above three methods, and their findings indicated that the LASSO outperforms FPC regression and penalized $\log$-likelihood in terms of misclassification rate and parameter estimation. However, the LASSO and penalized $\log$-likelihood methods may produce unstable estimates under two situations: when a large number of basis expansion functions are used to approximate the functional predictors or when a large number of functional predictors are used in the method \citep[see, for example][]{matsui2009, BS20}. Therefore, using dimension reduction techniques such as FPC regression is more often than using other methods like $\log$-likelihood and LASSO in scalar-on-function regression \citep{Reiss2017}.

The infinite-dimensional functions are projected into a finite-dimensional space of orthonormal bases in FPC regression. Then, the regression model of binary response on the projections of the functional predictors, called principal component scores, is used to approximate the regression model of binary response on the functional predictors. While doing so, the FPCs and the corresponding scores are computed based on the covariance between the functional predictors. However, a few FPCs generally comprise most of the covariance between the functional predictors. Thus, they may not necessarily be important to classification accuracy when representing the FPCs. All or some of the most important terms accounting for the interaction between the basis functions and functional predictors might come from later FPCs \citep[see, e.g.,][for more information]{Delaigle2012}. To overcome this problem, \cite{Escabias2007} proposed a functional partial least squares (FPLS) regression to estimate the FLogR model. FPLS regression uses both response and predictors when extracting the FPLS components, capturing most information with fewer terms. Thus, in scalar-on-function regression models, the FPLS regression is usually preferred to FPC regression \citep{Reiss2007, Delaigle2012, Yu2016}. In addition, the numerical analyses of \cite{Escabias2007} showed that the FPLS regression provides more accurate parameter function estimation to those of FPC regression but with a larger dimension reduction. Therefore in this paper, we consider the FPLS regression to estimate the FLogR model.

In all the studies mentioned above, the parameter function in the FLogR model is estimated via the least-squares (LS) type estimator. However, the LS estimator is severely affected by the presence of outliers. In such a case, the LS estimator produces biased estimates, leading to an increased probability of misclassification. Little attention has been paid to robustly estimating the parameter function of the FLogR model in the presence of outliers. We found only one study by \cite{DenhereBillor} that estimates the parameter function of the FLogR model robustly. By using the minimum covariance determinant method, \cite{DenhereBillor} computed the FPC scores. Then, the binary outcome and the computed scores are used to approximate the FLogR model, where the maximum likelihood estimator is used to estimate this model. However, this model may not have a full robust property since the maximum likelihood estimator is used in the approximate model. To the best of our knowledge, there is no approach to robustly compute its FPLS components and estimate the regression coefficient function for the FLogR model.

In this study, we propose a robust FPLS (RFPLS) method to robustly estimate the parameter function in the FLogR model and classify the binary response in the presence of outliers. In our proposed method, we first extend the partial least squares (PLS) algorithm proposed by \cite{DodgeWhittaker} to the functional data. Then, we consider the weighted likelihood-based generalized linear model of \cite{Fatemah2016} to extract the FPLS components of the FLogR model robustly. In addition, a weighted likelihood-based method is used to estimate the parameters of the approximated model. Our proposed method is robust to vertical outliers in the response variable and leverage points in the predictor variable. Our numerical analyses, which will be discussed in detail in Section~\ref{sec:results}, reveal that the proposed method produces an improved performance to its competitors in both estimations of parameter function and classification of the binary response variable in the presence of outliers. Our records also reveal that the proposed method produces competitive results when no outlier is present in the data.

The remainder of this paper is organized as follows. Section~\ref{sec:methodology} is devoted to introducing the scalar-on-function logistic regression. In Section~\ref{sec:RFPLS}, we present our robust FPLS method in the FLogR model. The finite-sample performance of the proposed method is evaluated via several Monte Carlo experiments as well as an empirical data set, and the results are presented in Section~\ref{sec:results}. Section~\ref{sec:conc} concludes the paper.

\section{Scalar-on-function logistic regression} \label{sec:methodology}

Let us consider an independent and identically distributed (i.i.d.) random sample $\{Y_i,~ \X_i(t):$ $i = 1, 2, \ldots, n\}$ from a population $\left \{ Y, \X \right \}$, where $Y \in \left \{ 0,1 \right \}$ is a binary outcome and $\X = \left \{ \X(t) \right \}_{t \in \mathcal{I}}$ is a stochastic process defined in a $\mathcal{L}_2$ separable Hilbert space with a bounded and closed interval $t \in \mathcal{I}$. The FLogR model used to explore the relationship between $Y$ and $\X(t)$ is defined as follows: \citep{Ratcliffe}:
\begin{equation}\label{eq:flr}
Y_{i}=\pi_{i}+\varepsilon_{i}, i=1,\ldots,n,  
\end{equation}
where $\varepsilon_{i}$ is the independent random error term with mean-zero and variance $\pi_{i}(1-\pi_{i})$ and $\pi_{i}$ is the probability that the response variable takes value one given an observation of the functional predictor
\begin{equation}\label{denk3}
\pi_{i} = P\left [Y=1|\X(t)=\X_{i}(t)  \right]  = \frac{\exp\left \{ \alpha + \int_{\mathcal{I}}\X_{i}(t)\beta(t)dt \right \}}{1+\exp\left \{ \alpha + \int_{\mathcal{I}}\X_{i}(t)\beta(t)dt \right \}},
\end{equation}
where $\alpha$ is the intercept parameter and $\beta(t)$ denotes the regression coefficient function. From~\eqref{denk3}, the logit transformation yields 
\begin{equation*}
l_{i}=\ln \left(\frac{\pi_{i}}{1-\pi_{i}}\right)= \ln\frac{P\left[Y=1|\X(t)=\X_{i}(t)\right]}{P\left[ Y=0|\X(t)=\X_{i}(t)\right]}.
\end{equation*}
In what follows, the FLogR model~\eqref{eq:flr} can be expressed in terms of logit transformation as follows:
\begin{equation}\label{denk5}
l_{i}=\alpha +\int_{\mathcal{I}}\X_{i}(t)\beta(t) dt.
\end{equation}

The expression~\eqref{denk5} renders possible to interpret the regression coefficient function so that the integral of $\beta(t)$ multiplied by a constant $\mu$ can be interpreted as the multiplicative change in the odds of response $Y = 1$ obtained when a functional observation is incremented constantly in $\mu$ units along $\mathcal{I}$ \citep{Escabias2005, Escabias2007}. However, the estimation of Model~\eqref{denk5} is a difficult task as the regression coefficient function $\beta(t)$ belongs to an infinite-dimensional space. Another problem related to the estimation of Model~\eqref{denk5} is that while the functional predictor $\X(t)$ belongs to an infinite-dimensional space, it is observed in a finite set of discrete points. As in other scalar-on-function regression models, the common practical approach to overcoming these problems is reconstructing the functional form of the functional predictor using a finite-dimensional basis expansion method. Some possible basis functions are polynomial basis functions (which are constructed from the monomials $\phi_k(t) = t^{k-1}$), Bernstein polynomial basis functions (which are constructed from $1, 1-t, t, (1-t)^2, 2t(1-t), t^2, \dots)$, Fourier basis functions (which are constructed from $1, \sin(wt), \cos(wt), \sin(2wt), \cos(2wt), \dots)$, radial basis functions, $B$-spline basis functions, wavelet basis functions and extracted basis functions from data (FPCs). Alternatively, one could also use a nonparametric smoothing technique to smooth or interpolate functions depending on the underlying behavior of the data \citep[see, e.g.,][for more information]{Ratcliffe, RamsaySilverman2005, Escabias2005, Amato2006, FerratyVieu, MousaviSorensen, KimKim2018}. This study considers the $B$-spline basis expansion method to reconstruct the functional form of the discretely observed functional predictor and estimate the infinite-dimensional regression coefficient function.

With a sufficiently large number of basis expansion functions $K$, the functional predictor $\X(t)$ can be approximated as a linear combination of the basis functions $\phi(t)$ and the corresponding basis expansion coefficients $a$ as follows:
\begin{equation}\label{eq:bex}
\X(t)\approx \sum_{k=1}^{K}a_{k}\phi_{k}(t) = \bm{a}^\top \bm{\phi}(t). 
\end{equation}
Similarly, the regression coefficient function can be expressed in the basis $\bm{\phi}(t)$ as follows:
\begin{equation}\label{eq:beb}
\beta(t)\approx \sum_{k=1}^{K}\beta_{k}\phi_{k}(t) = \bm{\beta}^\top \bm{\phi}(t).
\end{equation}
Substituting~\eqref{eq:bex} and~\eqref{eq:beb} into~\eqref{denk5}, the FLogR model in terms of logit transformation is expressed as follows:
\begin{equation*}
l_{i} = \alpha + \sum_{k=1}^K a_{ik}\phi_{k}(t) \beta_{k}\phi_{k}(t).
\end{equation*}
Let us denote by $\bm{\Psi} = \int_{\mathcal{I}}\bm{\phi}(t)\bm{\phi}^\top(t)dt$ the $K\times K$ inner product matrix of the basis functions. Then, the logit transformation expression of the FLogR model~\eqref{eq:flr} has the following matrix form:
\begin{equation*}
\bm{L} = \mathbf{1}\alpha + \bm{A} \bm{\Psi} \bm{\beta},
\end{equation*}
where $\bm{L} = \left [ l_{1}, \ldots,l_{n}\right ]^\top$, $\mathbf{1} = \left [1,1, \ldots,1\right ]^\top$, and $\bm{A}$ is the $n\times K$ matrix with row entries $a_{i}=\left [a_{i1},a_{i2}, \ldots,a_{iK}\right ]$.

The results presented above demonstrate that the infinite-dimensional FLogR model of $Y$ on $\X(t)$ can be reduced to a simple finite-dimensional logistic regression (LogR) model of $\bm{L}$ on the random matrix $\bm{A} \bm{\Psi}$. Let $\widehat{\beta}$ denote an estimate of $\beta$. Then, the regression coefficient function in~\eqref{denk5} can be approximated using the basis functions and the estimated basis expansion coefficient as follows:
\begin{equation*}
\widehat{\beta}(t) = \widehat{\bm{\beta}}^\top \bm{\phi}(t).
\end{equation*}

Due to the nature of functional data, the between columns of obtained design matrix $\bm{A} \bm{\Psi}$ become highly correlated (multicollinearity). In such a case, the estimate of $\widehat{\bm{\beta}}$ obtained via the LS method may have a large variance. While the FPLS method of \cite{Escabias2007} overcomes the multicollinearity problem, it may be significantly affected by outliers because the FPLS is based on the LS estimator. When outliers are present in the data, the effects of these points are included in the FPLS components approximated from the random matrix $\bm{A} \bm{\Psi}$. The effects of outliers are also included in the estimated regression coefficient function $\widehat{\beta}(t)$ because its basis expansion coefficient $\bm{\beta}$ is estimated from the LogR model of $\bm{L}$ on the extracted PLS components. Thus, in the presence of outliers, the FPLS method based on the LS type estimator may produce biased parameter estimates and incorrect classification results. Therefore in this paper, we propose an RFPLS method to robustly estimate the regression coefficient function $\beta(t)$ and correctly classify the values of the binary response variable in the presence of outliers.

\section{The RFPLS method}\label{sec:RFPLS}

Before introducing the proposed method, we start with a brief introduction of the weighted likelihood-based LogR model. The weighted likelihood estimator (WLE) is derived from the weighted likelihood estimation equations by modifying the weights, which are functions of properly defined residuals of the maximum likelihood estimation equations. The WLE of an unknown common parameter $ \theta $ for an $n$-dimensional random sample $\left \{ Y_{1},\ldots,Y_{n} \right \}$ is obtained as a solution of the following estimating equation:
\begin{equation*}
\sum_{i=1}^{n}w\left ( Y_{i};\theta ,f^{*} \right )u\left ( Y_{i},\theta \right )=0, 
\end{equation*}
where $u\left ( Y_{i},\theta \right )$ is the maximum likelihood score function from the hypothesized model and $w\left ( Y_{i};\theta ,f^{*} \right )$ denotes the weights. Weights reduce the effects of outliers in the data set and on the score equations. The weights are defined as follows:
\begin{equation*}
w\left ( Y;\theta,f^{*} \right ) = \max\left ( \min\left ( \frac{\mathcal{A}\left [ \delta \left ( Y;\theta,f^{*} \right ) \right ]+1}{\delta \left ( Y;\theta,f^{*} \right )+1},1 \right ),0 \right ),
\end{equation*}
where $ \delta \left ( Y;\theta,f^{*} \right )$ and $\mathcal{A}\left [ \delta \left ( \cdot  \right )\right ]$ denote the Pearson residual function and residual adjustment function (RAF), respectively. The Pearson residual function, which indicates the harmony between the hypothesized model and the empirical distribution of observations is defined by $\delta \left ( Y;\theta,f^{*} \right )=\frac{f^{*}\left ( Y \right )}{m^{*}\left ( Y;\theta  \right )}-1$. For the LogR model, $f^{*}\left(Y \right )$ is the proportion of sample observations with value $Y$ and $m^{*}\left ( Y;\theta  \right )$ is the corresponding probability under the hypothesized model. The RAF, on the other hand, is a strictly increasing and twice differentiable function defined in the range $\left [-1,\infty \right )$.

Let us consider $n$ independent binary responses $Y=\left \{ Y_{1},, \ldots, Y_{n} \right \}$ and the corresponding $n\times K$ design matrix $\bm{H} = \bm{A} \bm{\Psi}$. For the LogR model, the maximum likelihood estimator (MLE) is found by equating the score functions obtained by taking the derivative of the following log-likelihood function with respect to the parameters to zero:
\begin{equation*}
\ln L\left ( \bm{\beta}  \right )=\sum_{i=1}^{n}Y_{i} \ln[\pi \left (\bm{H}_{i} \right )]+\left ( 1-Y_{i} \right ) \ln\left [ 1-\pi\left ( \bm{H}_{i} \right ) \right ],
\end{equation*}
where $\pi\left ( \bm{H}_{i} \right ) = P\left ( Y=1 \right|\bm{H}=\bm{H}_{i} ) = \frac{\exp\left ( \bm{H}_{i}^\top\beta  \right )}{1+\exp\left ( \bm{H}_{i}^\top\beta  \right )}$ is the conditional probability of the outcome where $\bm{H}_{i}= \left [ 1, \bm{H}_{i1},\ldots, \bm{H}_{iK} \right ]^\top$ is the $i$-th row of the design matrix $\bm{H}$ and $\bm{\beta} =\left [ \beta_{0}, \beta_{1}, \ldots, \beta_{p} \right ]^\top$ denotes the parameter vector. The score equations are nonlinear in $\bm{\beta}$, and thus, the iterative reweighted least squares (IRLS) method is usually used to obtain the solutions \citep[see e.g.,][for more information]{JohnFox2016}. In IRLS, the adjusted binary response variable vector $\bm{Z}$ with $i$-th element $\bm{Z}_{i}=\frac{Y_{i}-\pi (\bm{H}_{i})}{\Omega_{i}}+\bm{H}_{i}^\top \bm{\beta}$ is regressed on the columns of design matrix $\bm{H}$ using the weight diagonal matrix $\bm{\Omega}$ with diagonal elements $\bm{\Omega}_{i}=\pi(\bm{H}_{i})\left [ 1-\pi (\bm{H}_{i}) \right ]$. Note that the weight values ($\bm{\Omega}_{i}$'s) are the functions of $\pi(\bm{H}_{i})$. Therefore, the method is iterative and the $\bm{\Omega}_{i}$'s are renewed at each step to achieve the following solution
\begin{equation*}
\bm{\widehat{\beta}}_{MLE}=\left ( \bm{H}^\top \bm{\Omega} \bm{H} \right )^{-1} \bm{H} \bm{\Omega} \bm{Z}. 
\end{equation*}

The WLE of $\bm{\beta}$ is obtained by adding the weight diagonal matrix $\bm{W}$ based on the Pearson residuals to the maximum likelihood estimation equations and its two-step solution as follows:
\begin{equation*}
\bm{\widehat{\beta}}_{WLE}=\left ( \bm{H}^\top \bm{\Omega} \bm{W} \bm{H} \right )^{-1} \bm{H} \bm{\Omega} \bm{W} \bm{Z}.
\end{equation*}
In the first step of the solution, the weight diagonal matrix $\bm{W}$ is evaluated, and in the second step, the IRLS is used with a fixed $\bm{W}$. In each iteration of the IRLS method, $\bm{W}$ is updated and added to IRLS. The algorithm terminates when the values of $\bm{W}$s obtained in the consecutive steps do not change. Note that for non-Gaussian response distributions in generalized linear models, the distribution of the Pearson residuals is often skewed. Therefore, the Anscombe residuals whose asymptotic distributions are as close to Gaussian distribution as possible under the hypothesized model are usually used as an alternative to a more general solution. For the LogR model with a binomial response distribution, the Anscombe residual is defined as:
\begin{equation*}
r_{A_{i}}=\sqrt{m_{i}}\left [ B\left ( Y_{i},\frac{2}{3},\frac{2}{3} \right )-B\left ( \pi\left ( \bm{H}_{i} \right ),\frac{2}{3},\frac{2}{3} \right ) \right ]\left [ \pi \left ( \bm{H}_{i} \right )\left ( 1- \pi \left ( \bm{H}_{i} \right ) \right ) \right ]^{-\frac{1}{6}},
\end{equation*}
where $m_{i}$ denotes the trial at $i$-th observation and $B\left ( z,a,b \right )=\int_{0}^{z}t^{a-1}\left ( 1-t \right )^{b-1}dt$ denotes the beta function. The Anscombe residuals can easily be calculated using the \Rlogo\ package ``wle'' \citep{wle}. For more information, see also \cite{Fatemah2016}.

\subsection{The RFPLS method for the FLogR model}

The PLS method of \cite{DodgeWhittaker} starts with centering and scaling the predictor matrix. However, we consider the $L_1$ median as defined by \cite{Serrneels} and the median absolute deviation for robustly centering and scaling the random matrix $\bm{H}$, respectively. In addition, we consider a weighted version of the random matrix $\bm{H}$ to reduce the effects of outliers in the basis expansion coefficients. Let us denote by $w_i$ the weight for the $i$-th observation as follows:
\begin{equation*}
w_{i} = \underset{k}{\text{median}} \lbrace w_{ik} \rbrace, \qquad i = 1, \ldots, n,\quad k = 1, \ldots, K,
\end{equation*}
where $w_{ik}$ is the weight for $i$-th residual corresponding to the $k$-th column. Then, similar to \cite{AA17}, we consider weighting the random matrix $\bm{H}$ by multiplying each row of it by the square root of weights. As proven by \cite{Claudio2002}, the weights obtained from the weighted likelihood-based model hold $\sup \vert w_i - 1 \vert \xrightarrow{p} 0$. This result indicates that the weighted and unweighted versions of the design matrix $\bm{H}$ will be similar when no outlier is present in the data, and the weighted likelihood method tends to perform similar to the maximum likelihood method. On the other hand, when outliers are present in the data, the weights downweigh the effects of outliers in the design matrix. In this case, the weighted likelihood method is expected to perform better than the maximum likelihood method. Let $\widetilde{\bm{H}}$ denote the robustly centered and scaled as well as weighted version of $\bm{H}$. Let also $\widetilde{\bm{H}}_j$ for $j=1,\ldots,K$ denote the columns of the design matrix. Then, our proposed RFPLS method for the FLogR model is presented as follows:
\begin{enumerate}
\item[1)] \underline{Calculation of the RFPLS components. For $l = 1, 2, \ldots$; repeat:}
\begin{enumerate}
    \item Fit the weighted likelihood based LogR model $Y/\widetilde{\bm{H}}_j$, $\left ( j=1, \ldots,K \right )$. Denote by $\widehat{\bm{\vartheta}}_l = \left [\widehat{\vartheta}_{l1}, \ldots, \widehat{\vartheta}_{lK} \right ]^\top$ the estimated slope parameters and $\bm{V}_l = \left [v_{l1}, \ldots, v_{lK} \right ]^\top$ the normalized vector of $\widehat{\bm{\vartheta}}_l$ so that $v_{lj} = \widehat{\vartheta}_{lj} / \Vert \widehat{\vartheta}_{l} \Vert$ where $\Vert \cdot \Vert$ denotes the Euclidean norm.
    \item Set equal to zero the coefficients  $v_{lj}$ which are not significant according to the Wald test. More precisely, the elements of $\bm{V}_l$ satisfying $\left | \widehat{\vartheta}_{lj}/\text{SE}\left ( \widehat{\vartheta}_{lj}  \right )  \right |\leq z_{\alpha/2}$ is deleted. Here, the $\text{SE}\left ( \widehat{\vartheta}_{lj}  \right )$ and $z_{\alpha/2}$ denote the the estimated standard deviation of $\widehat{\vartheta}_{lj}$ and $\left(\alpha/2\right)$-th quantile of the standard Gaussian distribution, respectively.
    \item Compute the $l$-th robust PLS component as $\bm{T}_l = v_{l1} \widetilde{\bm{H}}_1 + \cdots + v_{lK} \widetilde{\bm{H}}_K$.
    \item Adjust the robust PLS component $\bm{T}_l$ by replacing the columns of the design matrix $\widetilde{\bm{H}}$ by their residuals as follows:
\begin{equation*}
\bm{H}_{j}-\text{E}\left ( \bm{H}_{j}|\bm{T}_{l} \right )=\text{E}\left ( \bm{H}_{j} \right )+\text{Cov}\left ( \bm{H}_{j},\bm{T}_{l} \right )\text{Var}\left ( \bm{T}_{l} \right )^{-1}\left [ \bm{T}_{l}-\text{E}\left ( \bm{T}_{l} \right ) \right ]. 
\end{equation*}
\end{enumerate}
The algorithm stops when all the elements of $\bm{T}_l$ are not significant because none of the estimated slope parameters $\widehat{\vartheta}_{lj}$ is not significant. 

\item[2)] \underline{LogR fitting of $Y$ on the retained robust PLS components.} \\
Let $\bm{\Gamma}$ denote the matrix of robust logit PLS components extracted from the design matrix $\widetilde{\bm{H}}$, i.e., $\bm{\Gamma} = \widetilde{\bm{H}} \bm{V}$, with $\bm{V}$ being the the matrix with column entries the coefficients of the robust logit PLS components in terms of original predictors. Then, the logit model in terms of the robust PLS components has the following expression:
\begin{equation*}
\widehat{\bm{L}} = \bm{1} \widehat{\alpha} + \bm{\Gamma} \widehat{\bm{\vartheta}},
\end{equation*}
where $\widehat{\bm{\vartheta}} = \left [ \widehat{\vartheta}_1, \ldots, \widetilde{\vartheta}_l \right ]^\top$ is the vector of WLEs of the coefficients of the logit model in terms of the $l$ robust PLS components obtained in the first step.

\item[3)] \underline{Robust estimation of the parameter function.} \\
The robust PLS regression model in terms of the original predictors can be expressed as follows:
\begin{equation*}
\widehat{\bm{L}} = \bm{1} \widehat{\alpha} + \widetilde{\bm{H}} \bm{V} \widehat{\bm{\vartheta}}.
\end{equation*}
Accordingly, the basis expansion coefficient of the regression coefficient function can be obtained as $\widehat{\bm{\beta}} = \bm{V} \widehat{\bm{\vartheta}}$. Finally, the robust estimation of the regression coefficient function is obtained as $\widehat{\beta}(t) = \widehat{\bm{\beta}}^\top \bm{\phi}(t)$.
\end{enumerate}

In summary, in our proposed RFPLS method, the effects of outliers in the functional predictors are downweighed by weighting the basis expansion coefficients (i.e., by weighting the design matrix $\bm{H}$) and computing the PLS components via the weighted likelihood method. In addition, in the last step of the proposed method, the basis expansion coefficients of the regression coefficient function are robustly obtained via the weighted likelihood method. Therefore, our proposed method is robust to both vertical outliers in the response variable and leverage points in the predictor variable.

\section{Numerical results} \label{sec:results}

We perform a series of Monte Carlo experiments under different data generation and outlier generation processes and empirical data analysis --- strawberry puree data, to investigate the finite-sample performance of the proposed RFPLS method. In our numerical analyses, we compare the empirical performance of the proposed method with FPLS and FPC. Throughout the numerical analyses, we note that four components are used to estimate the FPC regression models to ensure that at least 99\% of the variation in the functional predictors is captured. An example \Rlogo\ code for all the methods in this study can be found at \url{https://github.com/MugeMutis/RFPLS_FLogR}.

\subsection{Monte Carlo experiments}

\begin{figure}[!htb]
  \centering
  \includegraphics[width=8.7cm]{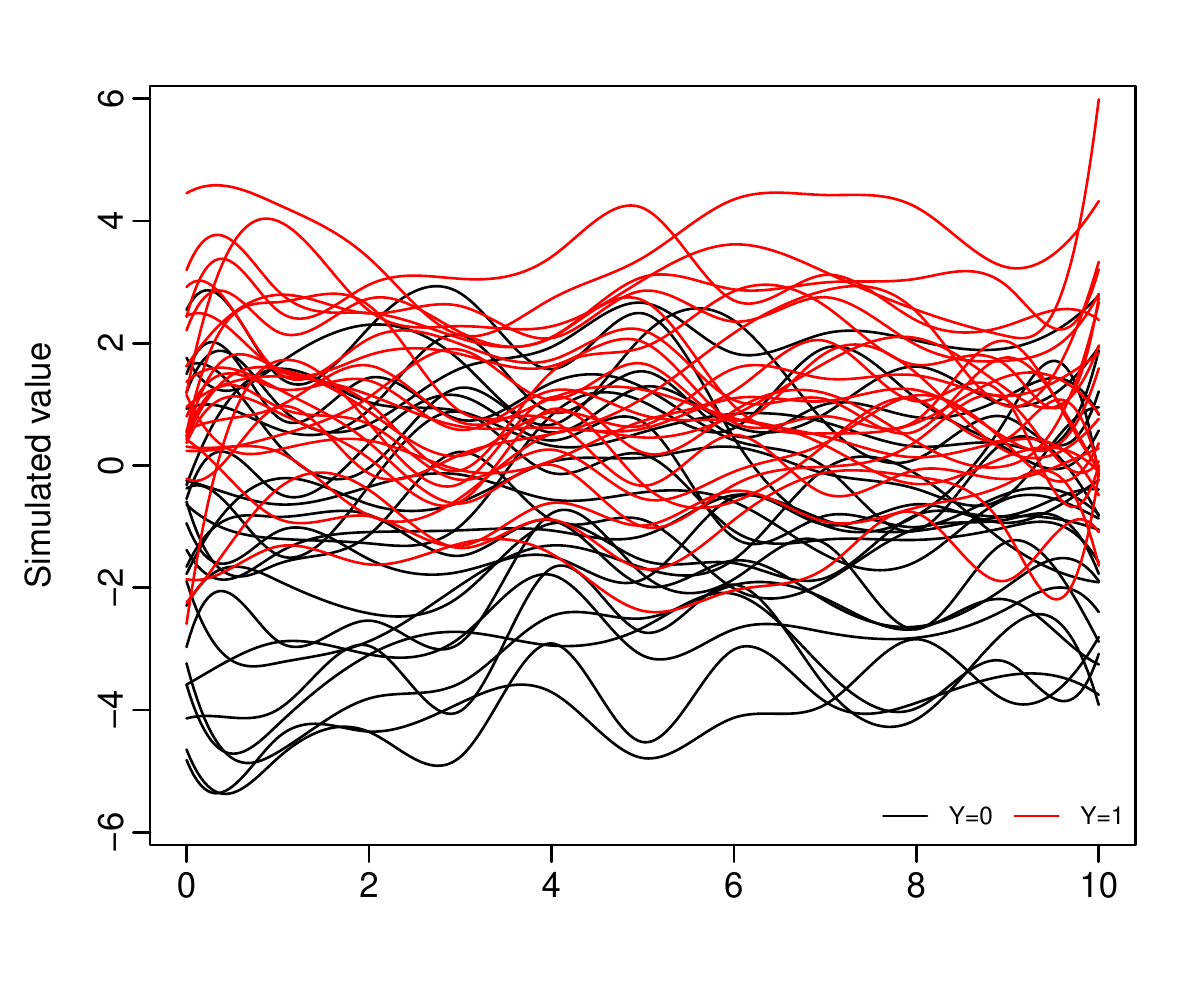}
\quad
  \includegraphics[width=8.7cm]{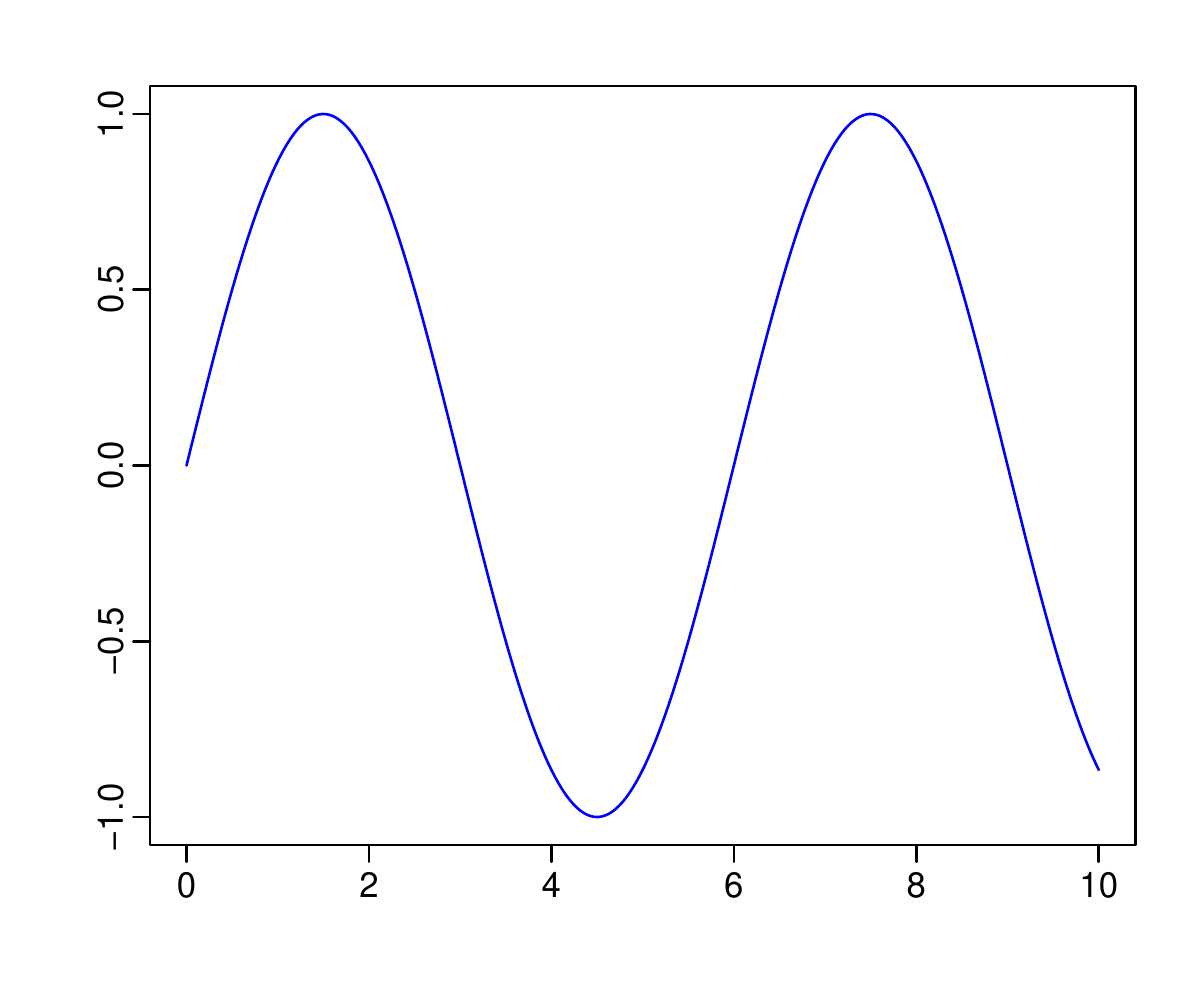}
\\  
  \includegraphics[width=8.7cm]{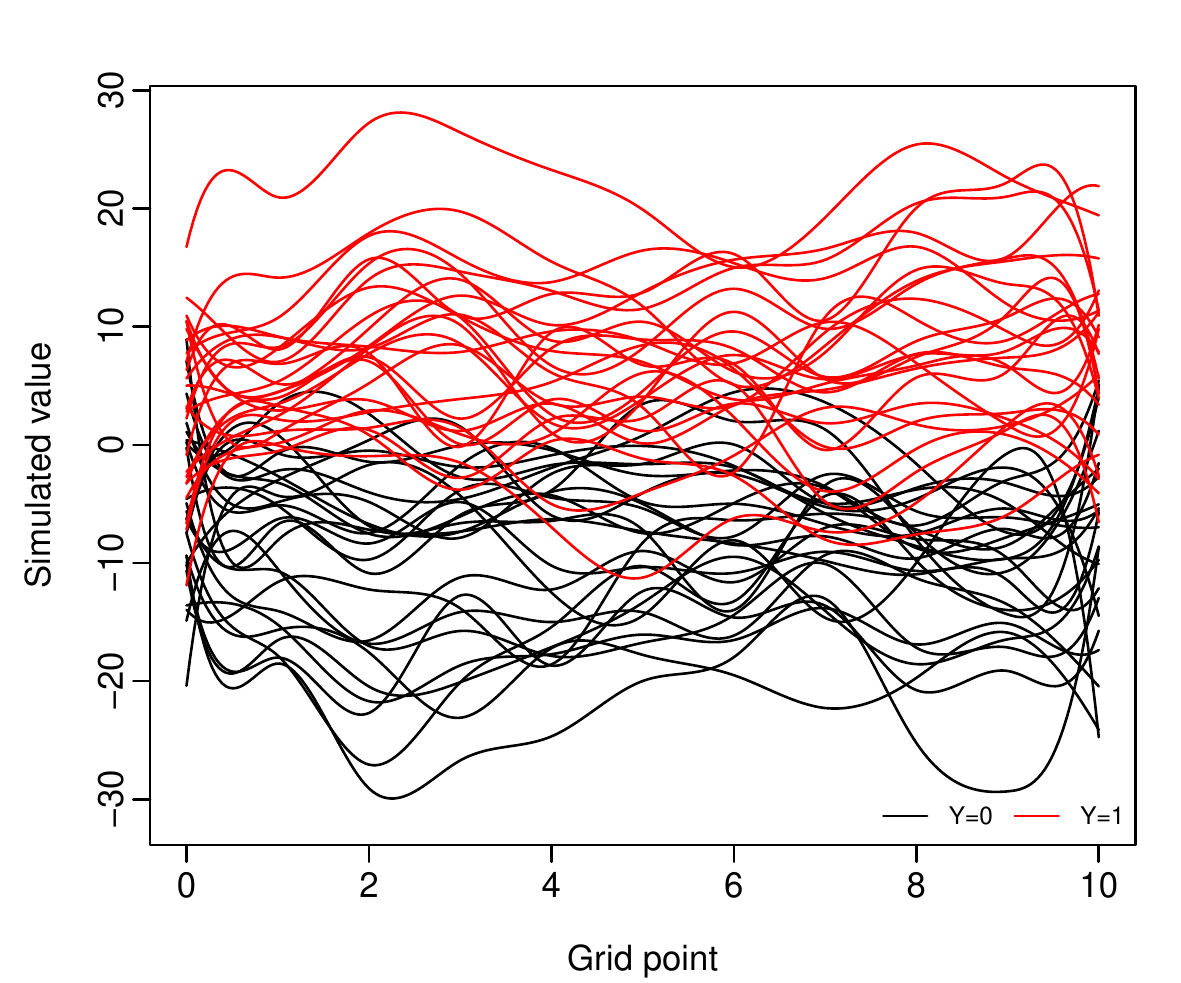}
\quad
  \includegraphics[width=8.7cm]{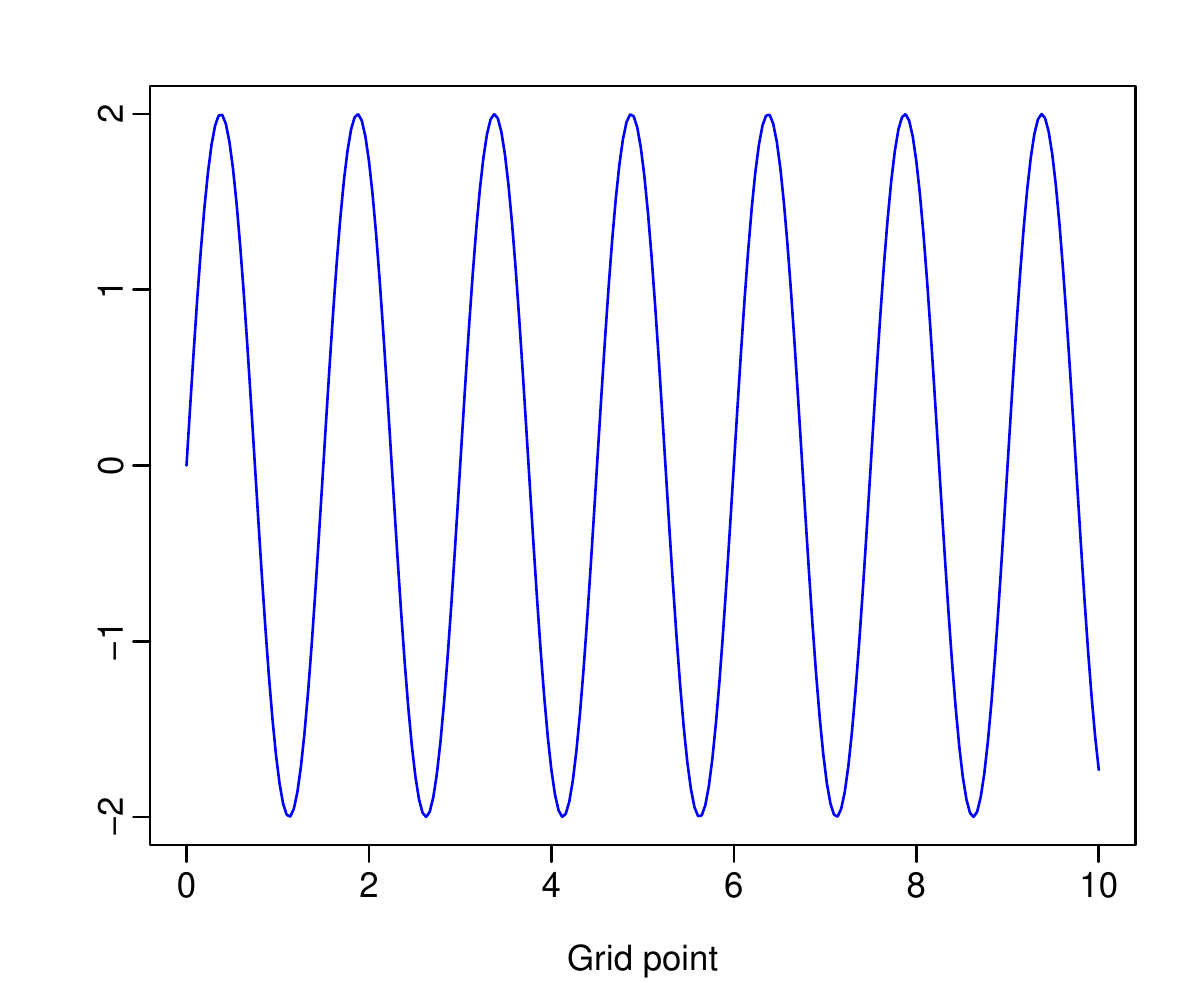}
  \caption{\small{Plots of the generated 25 sample curves for both $Y = 0$ and $Y = 1$ (left panels; scenario 1 (first row) and scenario 2 (second row)) and actual generated regression coefficient function (right panels; scenario 1 (first row) and scenario 2 (second row))}.}
  \label{fig:Fig_1}
\end{figure}

Throughout the Monte Carlo experiments, we consider two different data generation processes as follows:
\begin{description}
\item[Case 1.] In this case, we consider the data generation process of \cite{Mousavi2018}. First, we generate the trajectories of the functional predictor $\X(t)$ at 256 equally spaced point in the interval $\mathcal{I} = [0, 10]$. While doing so, 13 cubic $B$-spline basis functions $\left\{ \phi_k(t) \right\}_{k=1}^{13}$ are generated at nine equally spaced knots over the interval $[0, 10]$. In addition, the random basis expansion coefficients are generated as $\bm{C} = \bm{Z} U$, where $\bm{Z}$ is a $500 \times 13$ matrix of random values from the standard Gaussian distribution and $\bm{U}$ is a $13 \times 13$ random matrix consisting of values generated from the uniform distribution on $[0, 1]$. Then, the trajectories of the functional predictor is generated as $\X_i(t_j) = \sum_{k=1}^{13} c_{ik} \phi_k(t_j)$ where $j \in [0, 10]$. Finally, the values of the binary response are generated as a linear functional of the functional predictor and parameter function as follows:
\begin{equation*}
l_i = \int_0^{10} \X_i(t) \beta(t) dt,
\end{equation*}
where $\beta(t) = \sin \left( \frac{t \pi}{3} \right)$ and $t \in [0, 10]$.

In this case, we consider two scenarios. The data set is generated from the smooth process as presented above in the first scenario. In this scenario, we aim to confirm the correctness of the proposed method so that the proposed RFPLS method is expected to perform similarly to its traditional competitors. In the second scenario, the randomly selected $n \times [1\%,~5\%,~10\%]$ of the generated data are contaminated by outliers. In doing so, the randomly selected $n \times [1\%,~5\%,~10\%]$ trajectories from each class of the binary response of the functional predictor are generated similar to the first scenario but using $\bm{U} \sim$ uniform on $[0, 5]$ and $\beta(t) = 2 \sin \left( \frac{4 t \pi}{3} \right)$ where $t \in [0, 10]$. In addition, the corresponding binary response values are replaced by opposite classes so that the response variable contains outliers. Thus, we investigate the proposed method's performance against vertical outliers besides that leverage points. In this case, the proposed RFPLS method is expected to produce improved empirical performance over the classical methods. A graphical display of the data generated under this case is presented in Figure~\ref{fig:Fig_1}.
\end{description}

\begin{figure}[!htb]
  \centering
  \includegraphics[width=8.9cm]{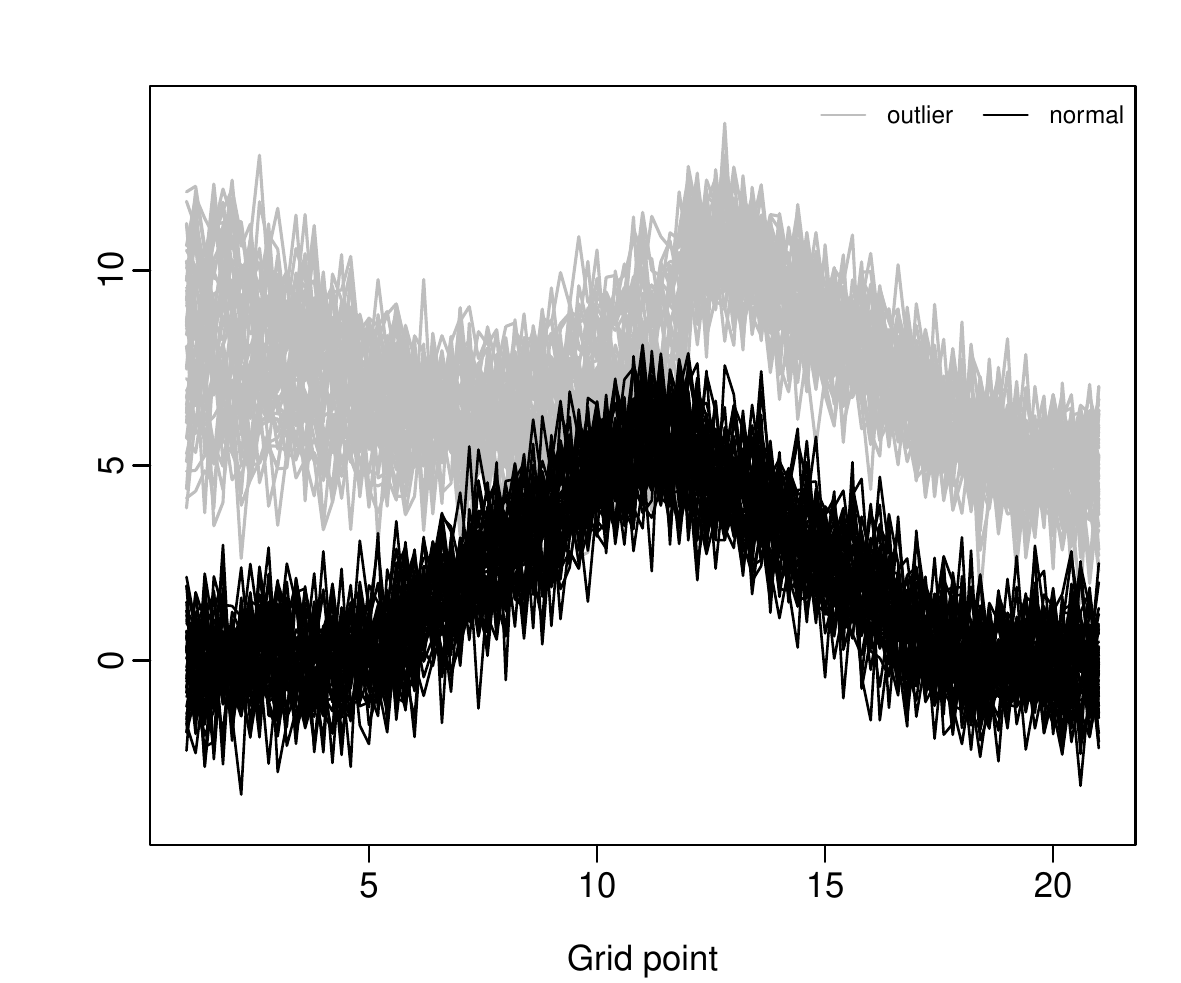}
  \includegraphics[width=8.9cm]{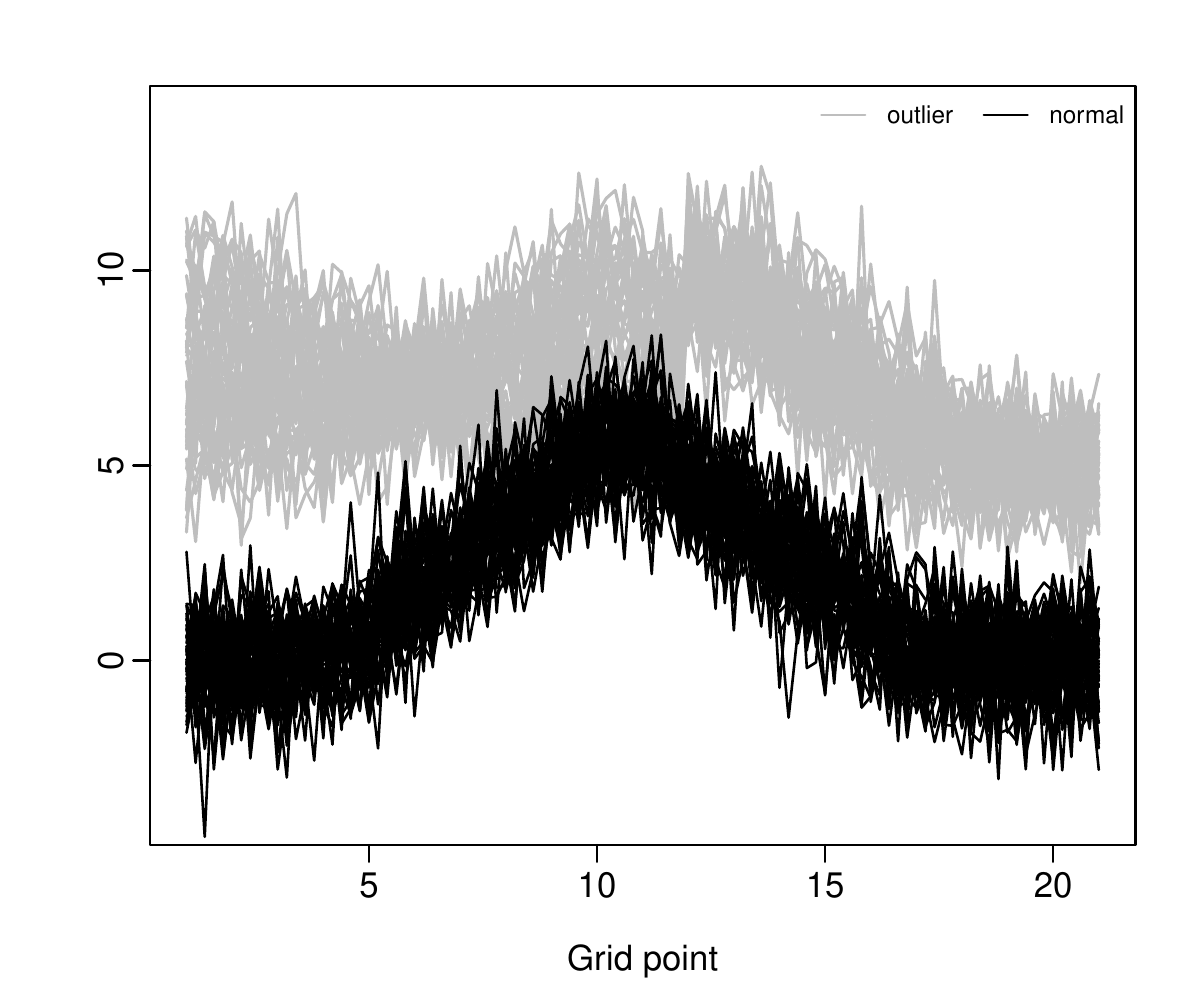}
  \caption{\small{Plots of the generated 50 sample curves for $Y = 0$ (left panel) and $Y = 1$ (right panel). The curves generated under scenario one are given in black, while those generated under scenario two are given in gray.}}
  \label{fig:Fig_2}
\end{figure}

\begin{description}
\item[Case 2.] In the second case, we focus on the curve discrimination accuracy of the FLogR model as in \cite{Escabias2007}. In this case, two different functional predictors are generated at 101 equally spaced point in the interval $\mathcal{I} \in [1, 21]$ based on the classes of the binary response, i.e., $Y = 0$ or $Y = 1$. For the first class ($Y = 0$), the trajectories of the functional predictor are generated as follows:
\begin{equation*}
\X(t) = u h_1(t) + (1-u) h_2(t) + \varepsilon(t),
\end{equation*}
where $u$ is a random variable generated from the uniform distribution on $[0, 1]$, $\varepsilon(t)$ is a random error generated from the standard Gaussian distribution, $h_1(t) = \max \left\{6 - \vert t - 11 \vert , 0 \right\}$, and $h_2(t) = h_1(t-4)$. For the second class ($Y = 1$), on the other hand, the trajectories of the functional predictor are generated as follows:
\begin{equation*}
\X(t) = u h_1(t) + (1-u) h_3(t) + \varepsilon(t),
\end{equation*}
where $h_3(t) = h_1(t+4)$.

Similar to \textbf{Case 1}, we also consider two scenarios in this case. In the first scenario, the data set is generated as presented above. In the second scenario, the randomly selected $n \times [1\%,~5\%,~10\%]$ trajectories of the functional predictor are generated as follows:
\begin{align*}
\X(t) &= u h_4(t) + (1-u) h_2(t) + \varepsilon^{*}(t), \qquad (\text{when}~Y = 0), \\
\X(t) &= u h_4(t) + (1-u) h_3(t) + \varepsilon^{*}(t), \qquad (\text{when}~Y = 1),
\end{align*}
where $\varepsilon^{*}(t)\sim \mathcal{N}(5, 1)$ and $h_4(t) = h_1(t + 50)$. In addition, the randomly selected $n \times [1\%,~5\%,~10\%]$ of the binary response values are replaced by opposite class. A graphical display of the data generated under this case is presented in Figure~\ref{fig:Fig_2}.
\end{description}

We generate $n = 1000$ trajectories for the functional predictor for each case and scenario. The following procedure is repeated 500 times to compare the empirical performance of the proposed RFPLS and traditional FPLS and FPC methods. First, the generated data set is randomly divided into training that include outliers and test samples with sizes 700 and 300, respectively. In each replication, the correct classification ratio (CCR) by taking 0.5 as a cut-point and the area under the receiver operating characteristic curve by considering all the possible cut-points (AUC) are computed to evaluate the classification performance among the methods. In addition, the integrated mean squared error (IMSE) is calculated under \textbf{Case 1} to compare the estimation performance of the methods as follows:
\begin{equation*}
\text{IMSE} = \int_{\mathcal{I}} \left[\beta(t) - \widehat{\beta}(t)\right]^2 dt.
\end{equation*}
We note that the data generated without outliers are first divided into the training and test samples in the Monte Carlo experiments. Then, only the randomly selected observations in the training samples are replaced by outliers. We also note that $K = 15$ basis expansion functions are used in the estimation phase of the methods.

\begin{figure}[!htb]
  \centering
  \includegraphics[width=8.9cm]{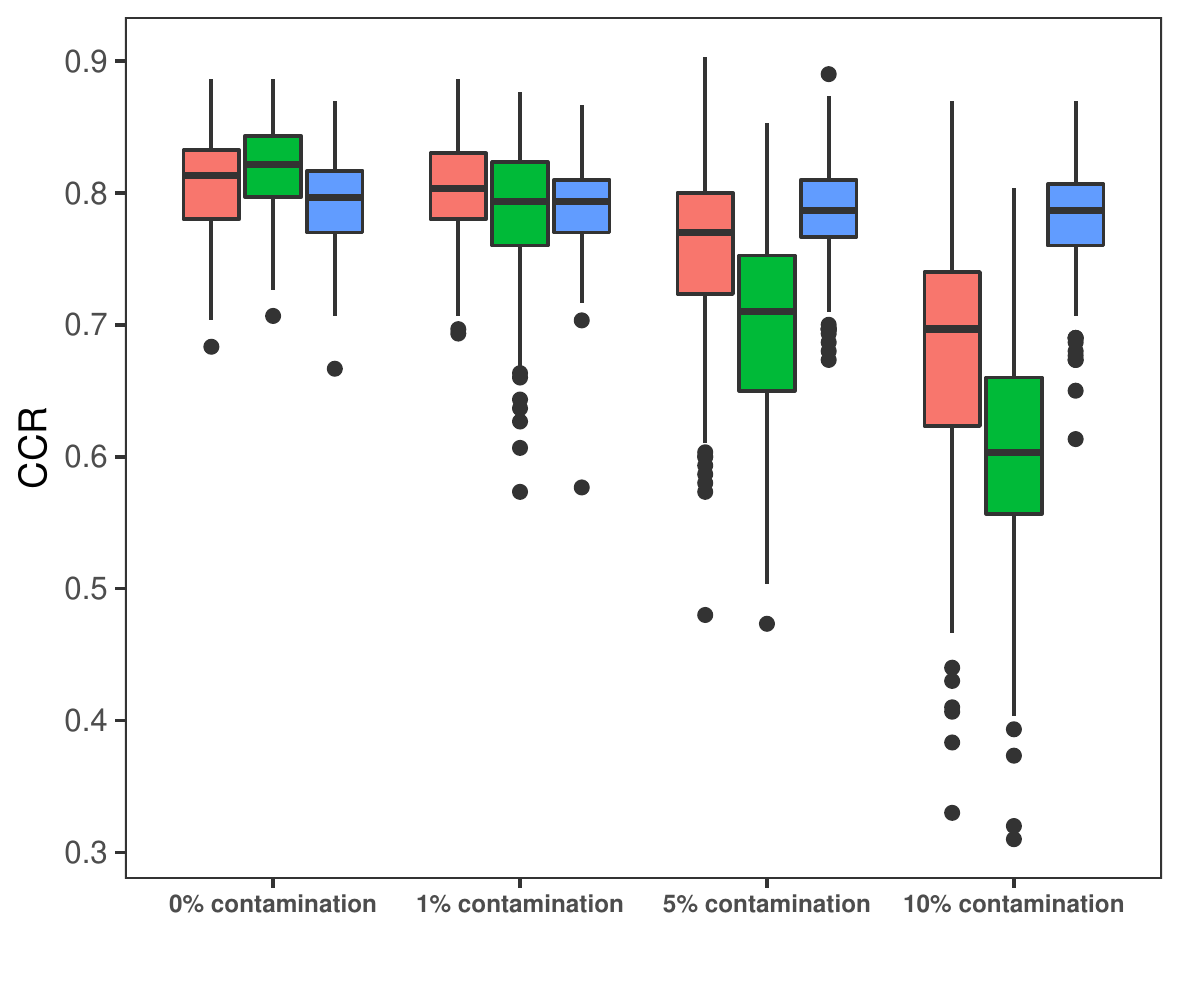}
  \includegraphics[width=8.9cm]{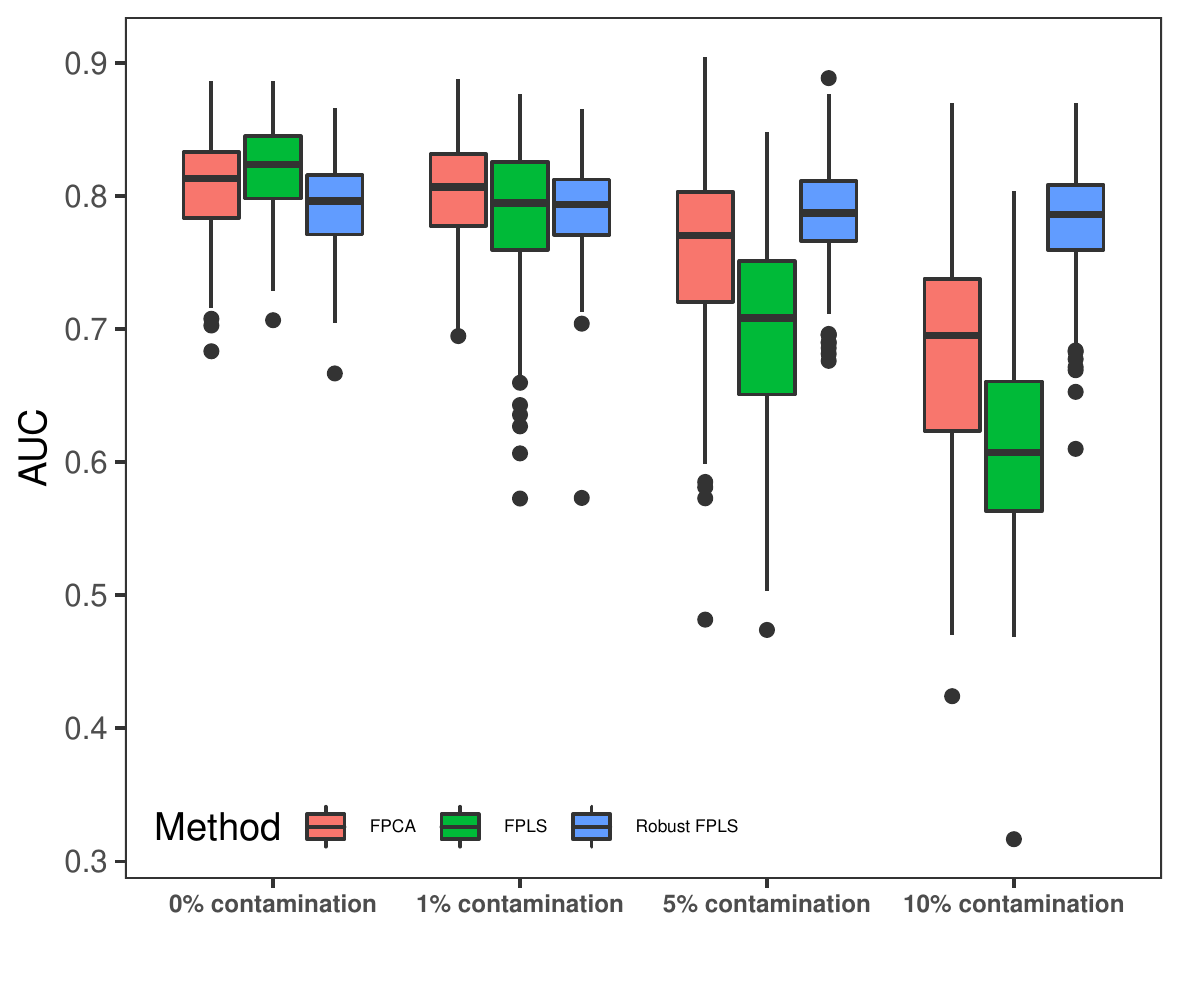}
  \caption{\small{Boxplots of the computed CCR (left panel) and AUC (right panel) values under Case 1 for the FPC, FPLS, and RFPLS methods}.}
  \label{fig:Fig_3}
\end{figure}

\begin{figure}[!htb]
  \centering
  \includegraphics[width=11cm]{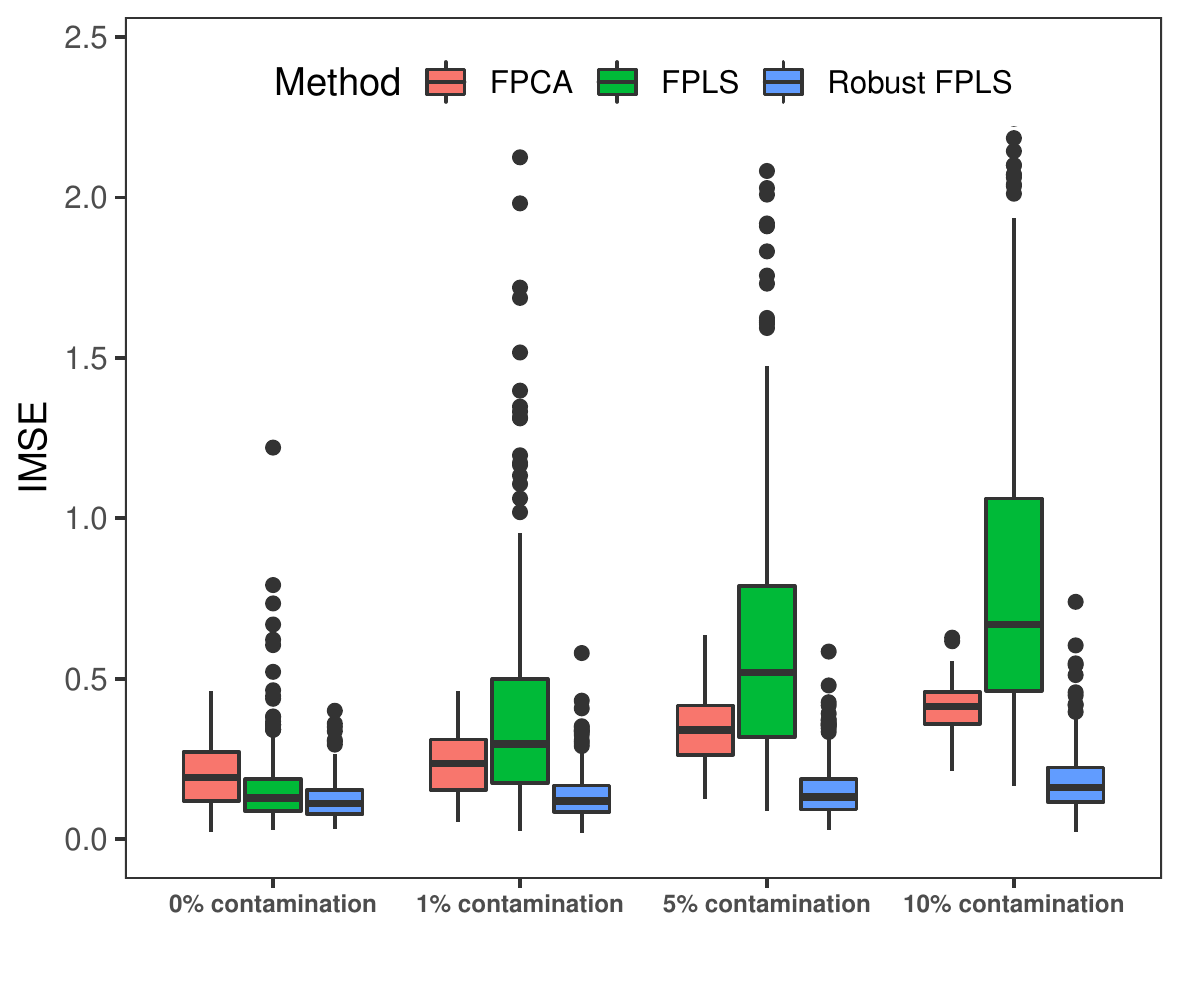}
  \caption{\small{Boxplots of the computed IMSE values under Case 1 for the FPC, FPLS, and RFPLS methods}.}
  \label{fig:Fig_4}
\end{figure}

\begin{figure}[!htb]
  \centering
  \includegraphics[width=8.9cm]{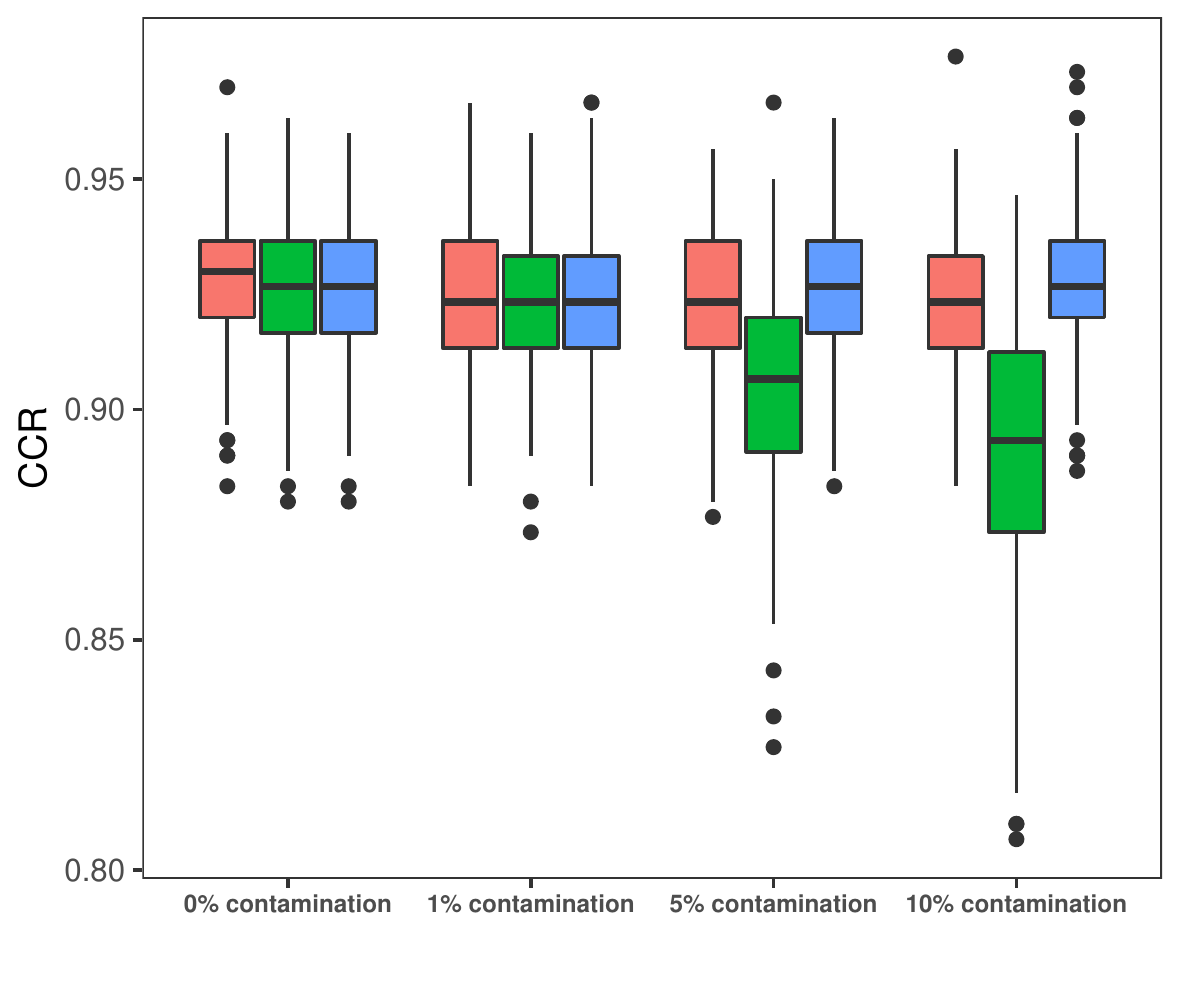}
  \includegraphics[width=8.9cm]{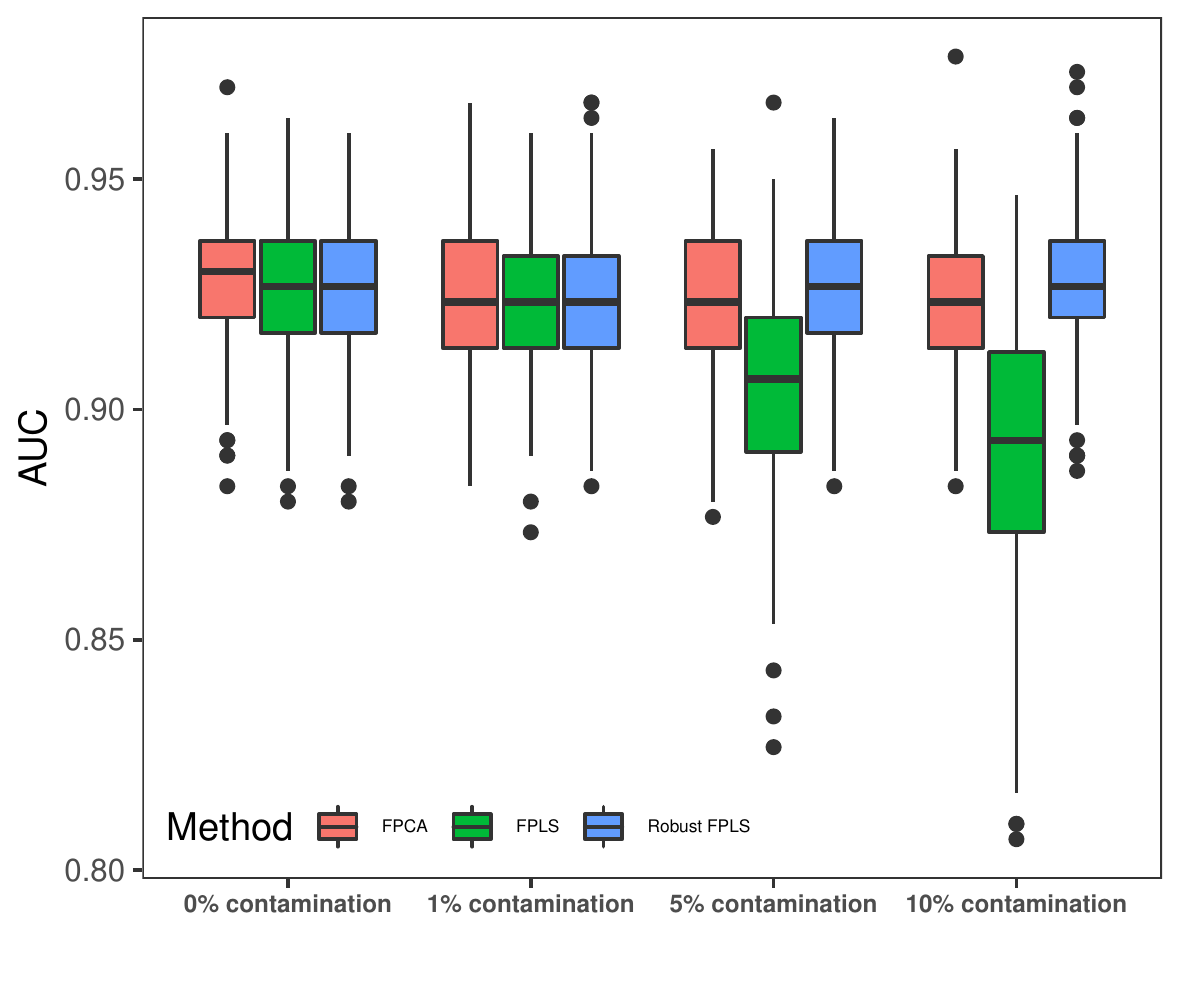}
  \caption{\small{Boxplots of the computed CCR (left panel) and AUC (right panel) values under Case 2 for the FPC, FPLS, and RFPLS methods}.}
  \label{fig:Fig_6}
\end{figure}

\begin{figure}[!htb]
  \centering
  \includegraphics[width=5.9cm]{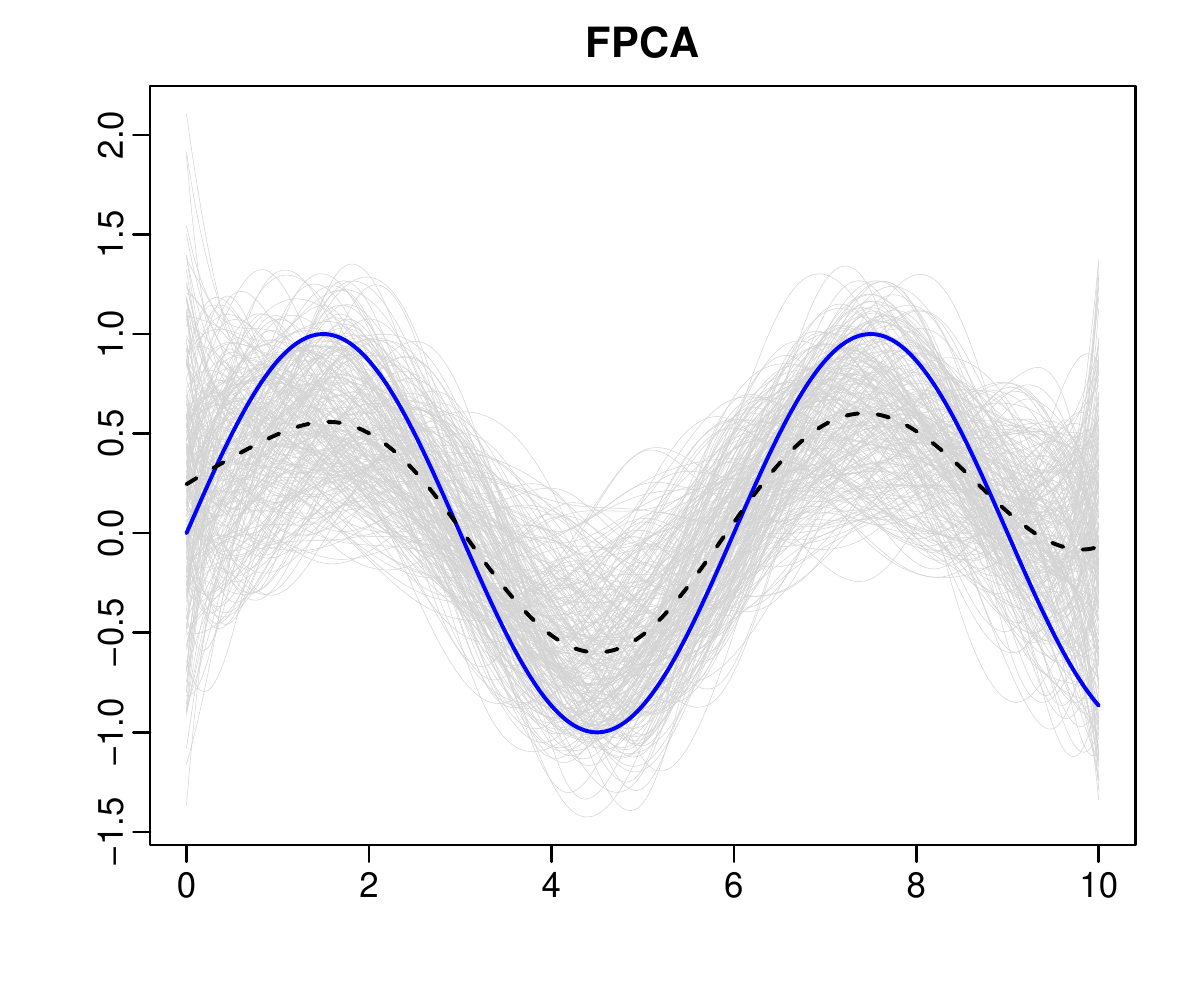}
  \includegraphics[width=5.9cm]{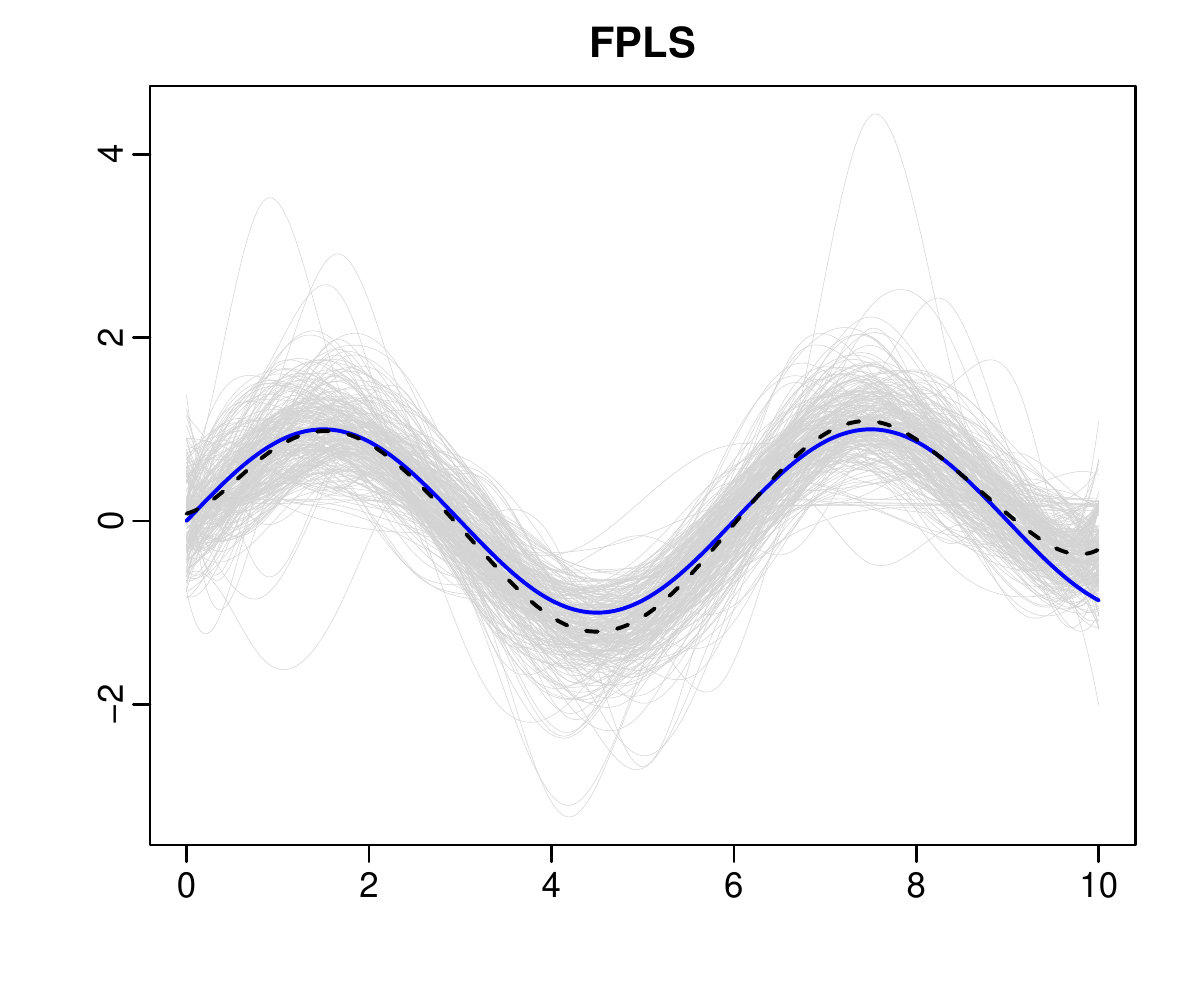}
  \includegraphics[width=5.9cm]{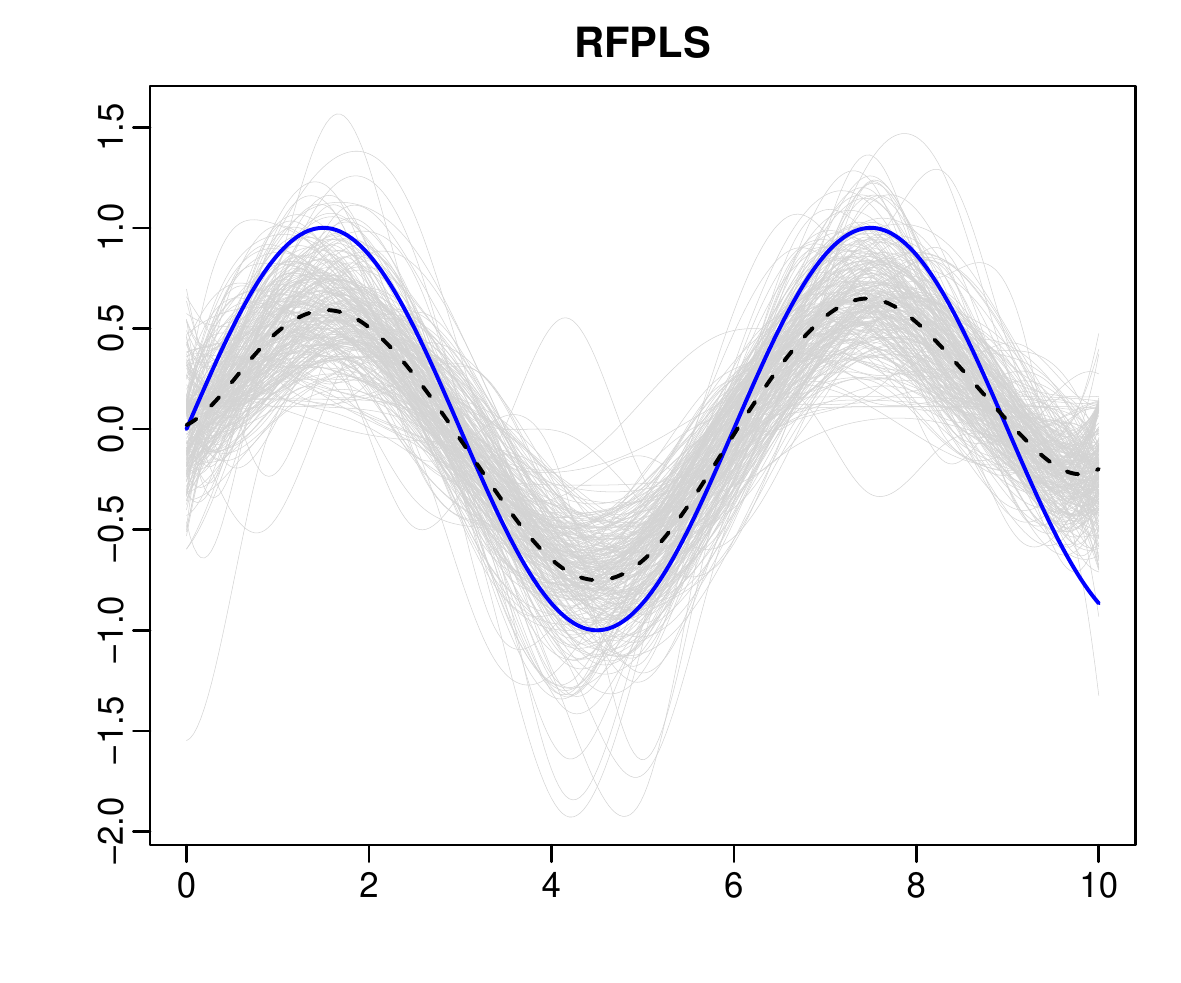}
\\  
  \includegraphics[width=5.9cm]{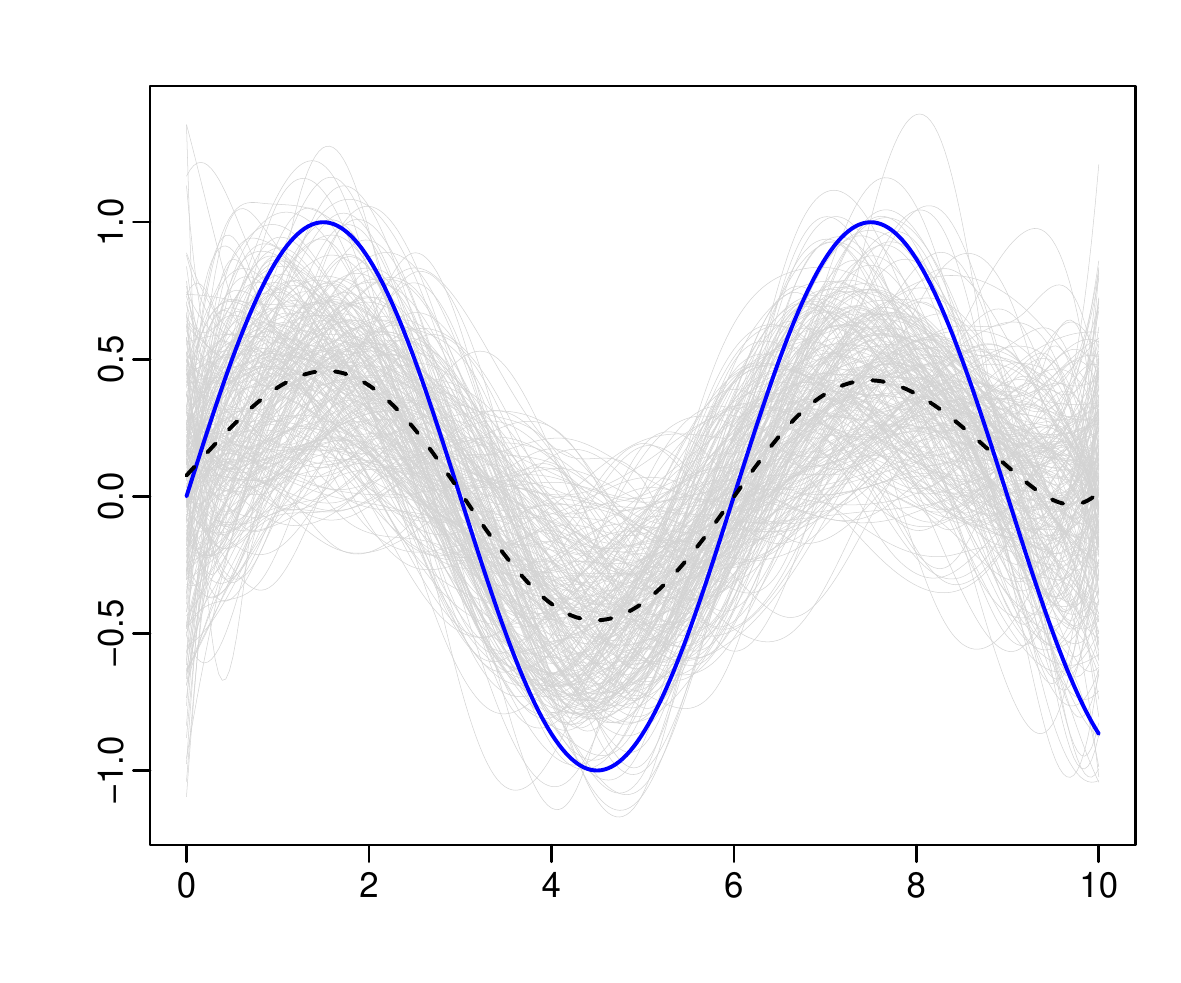}
  \includegraphics[width=5.9cm]{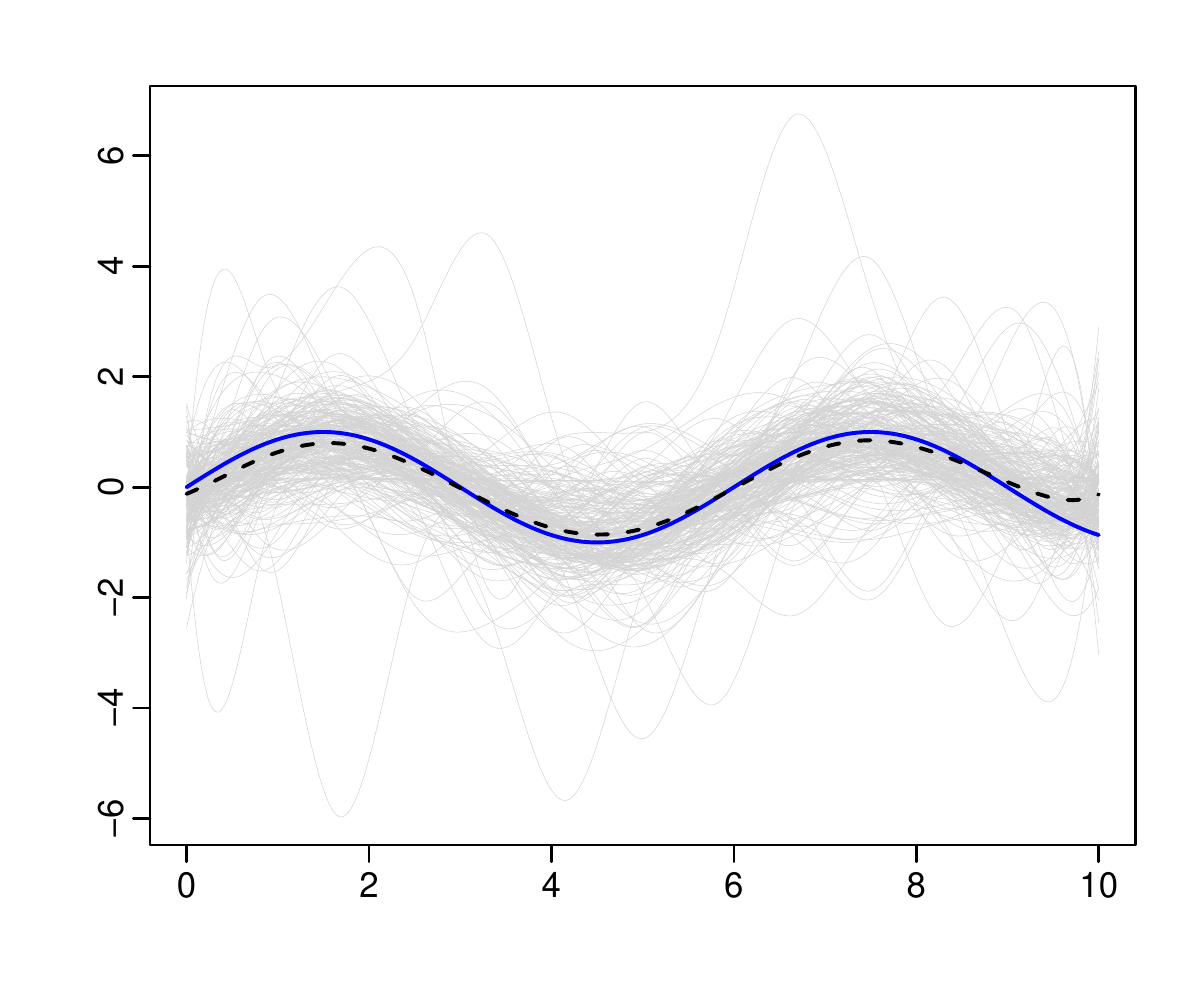}
  \includegraphics[width=5.9cm]{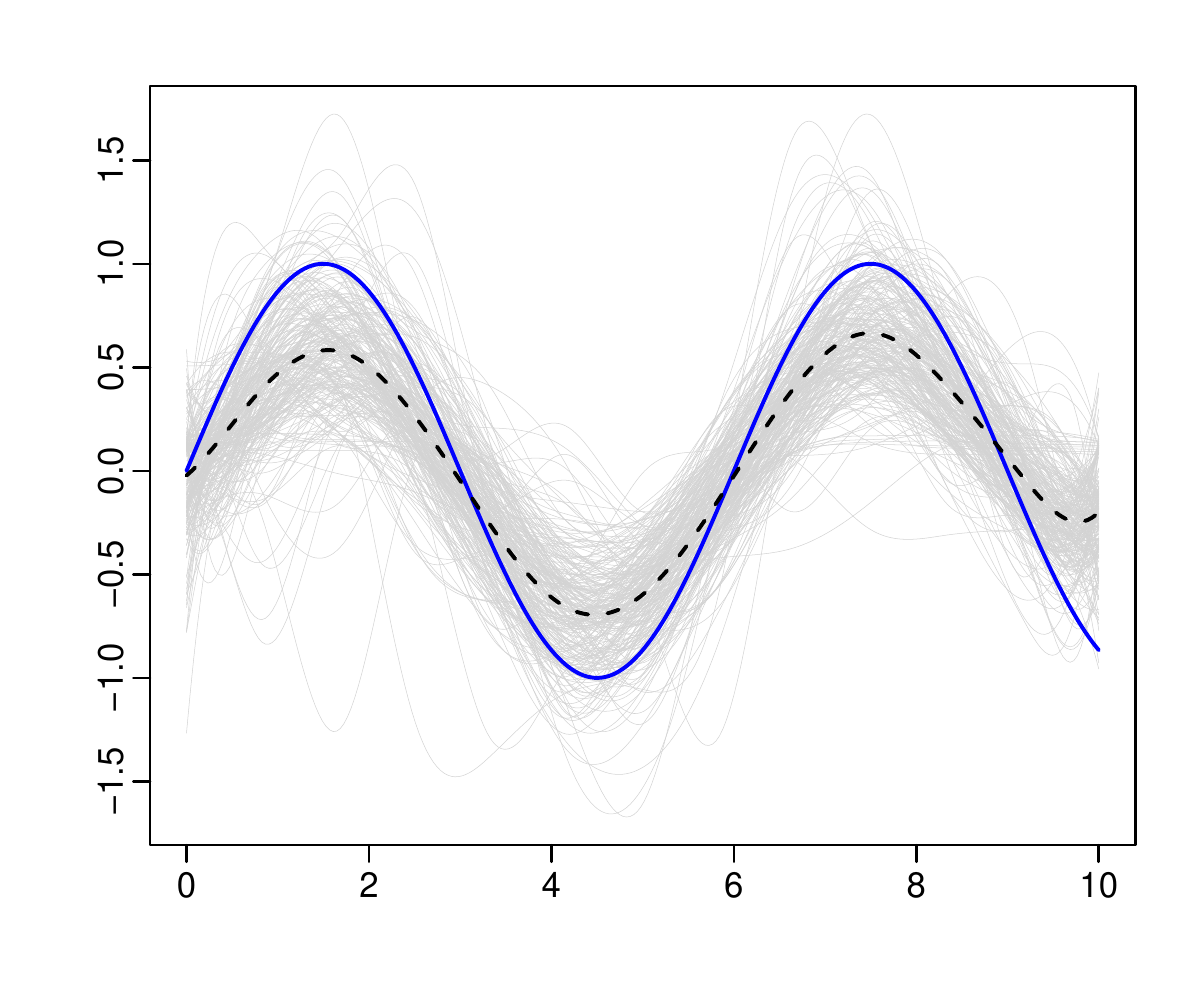}
\\
  \includegraphics[width=5.9cm]{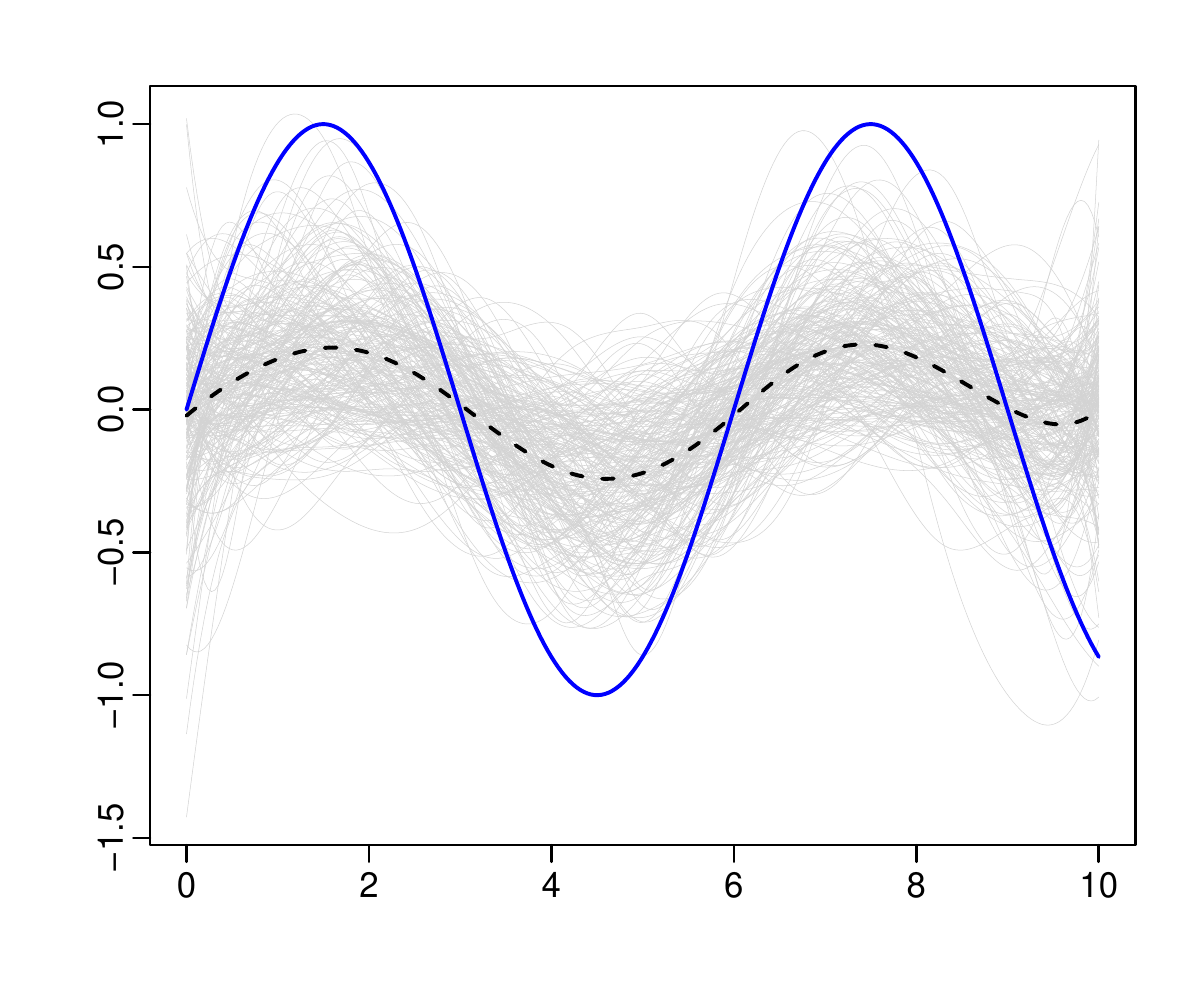}
  \includegraphics[width=5.9cm]{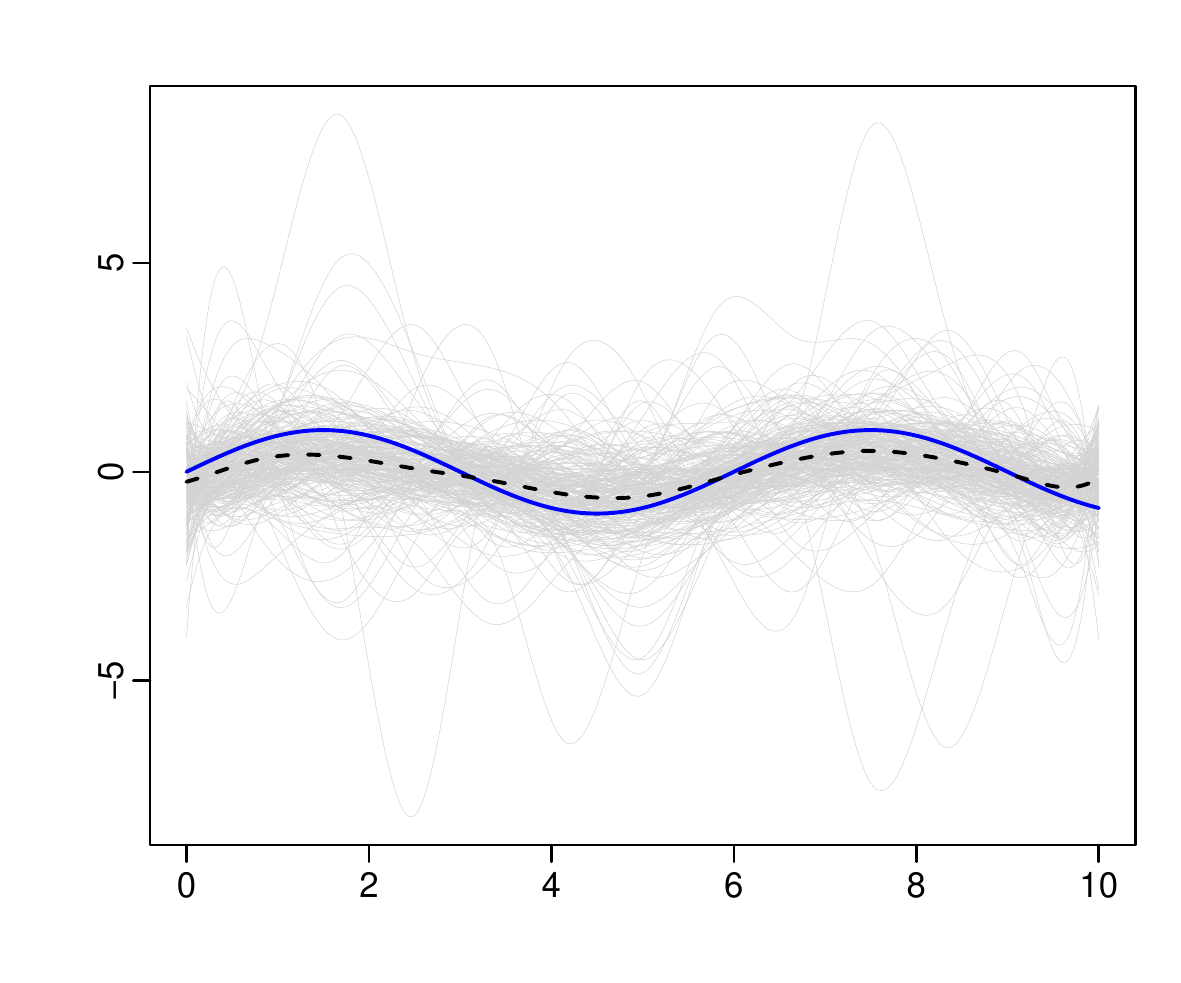}
  \includegraphics[width=5.9cm]{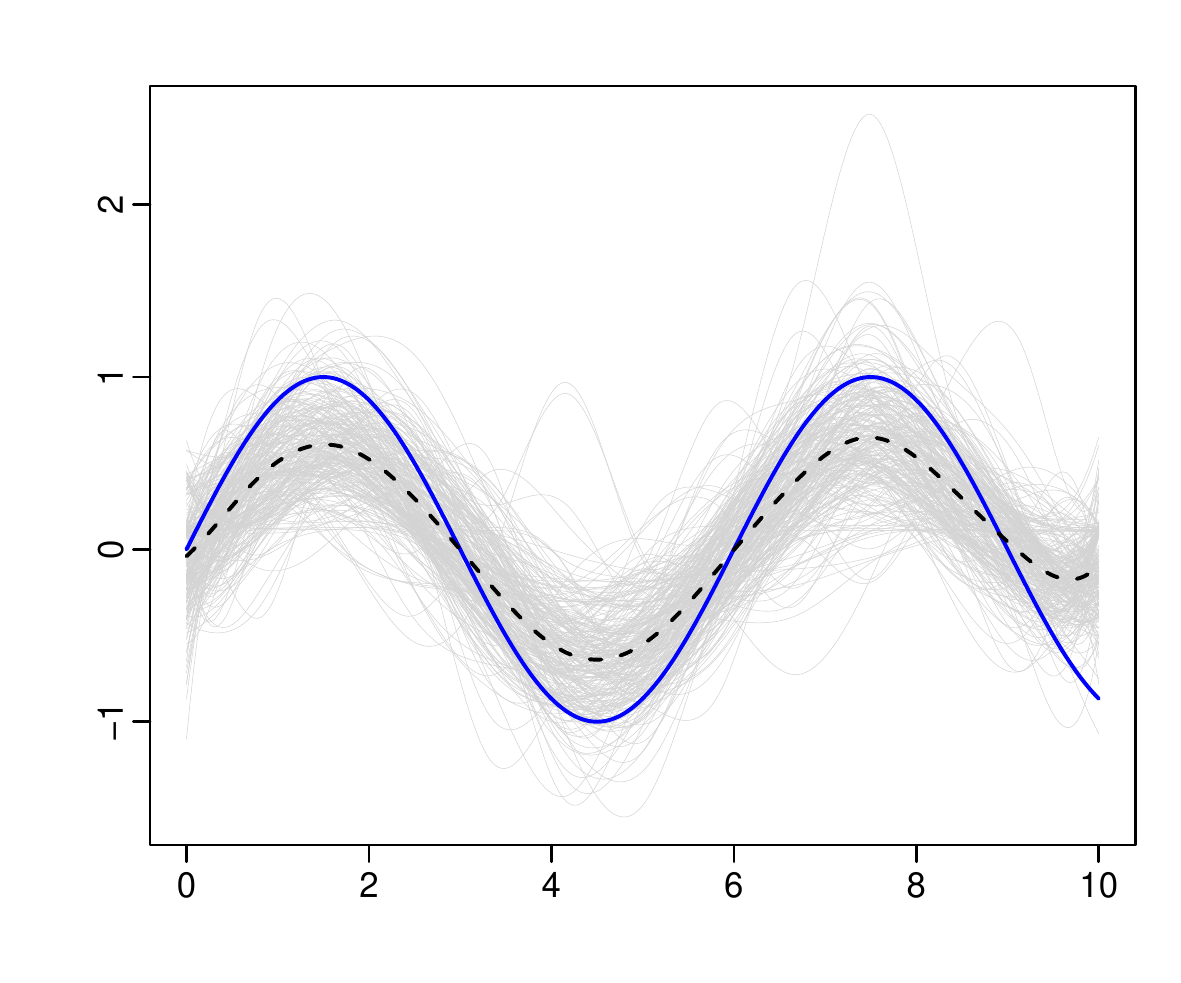}
  \\
  \includegraphics[width=5.9cm]{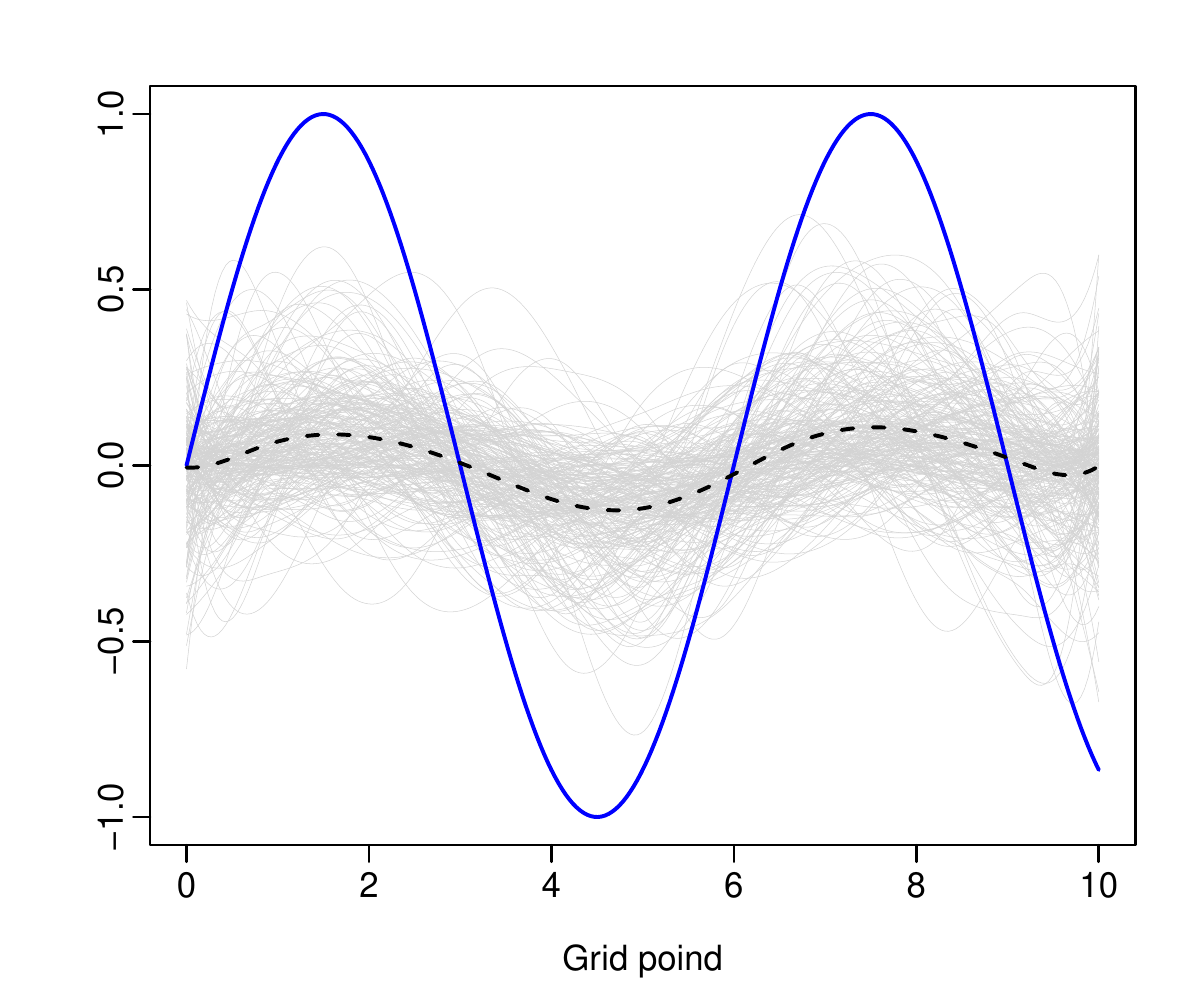}
  \includegraphics[width=5.9cm]{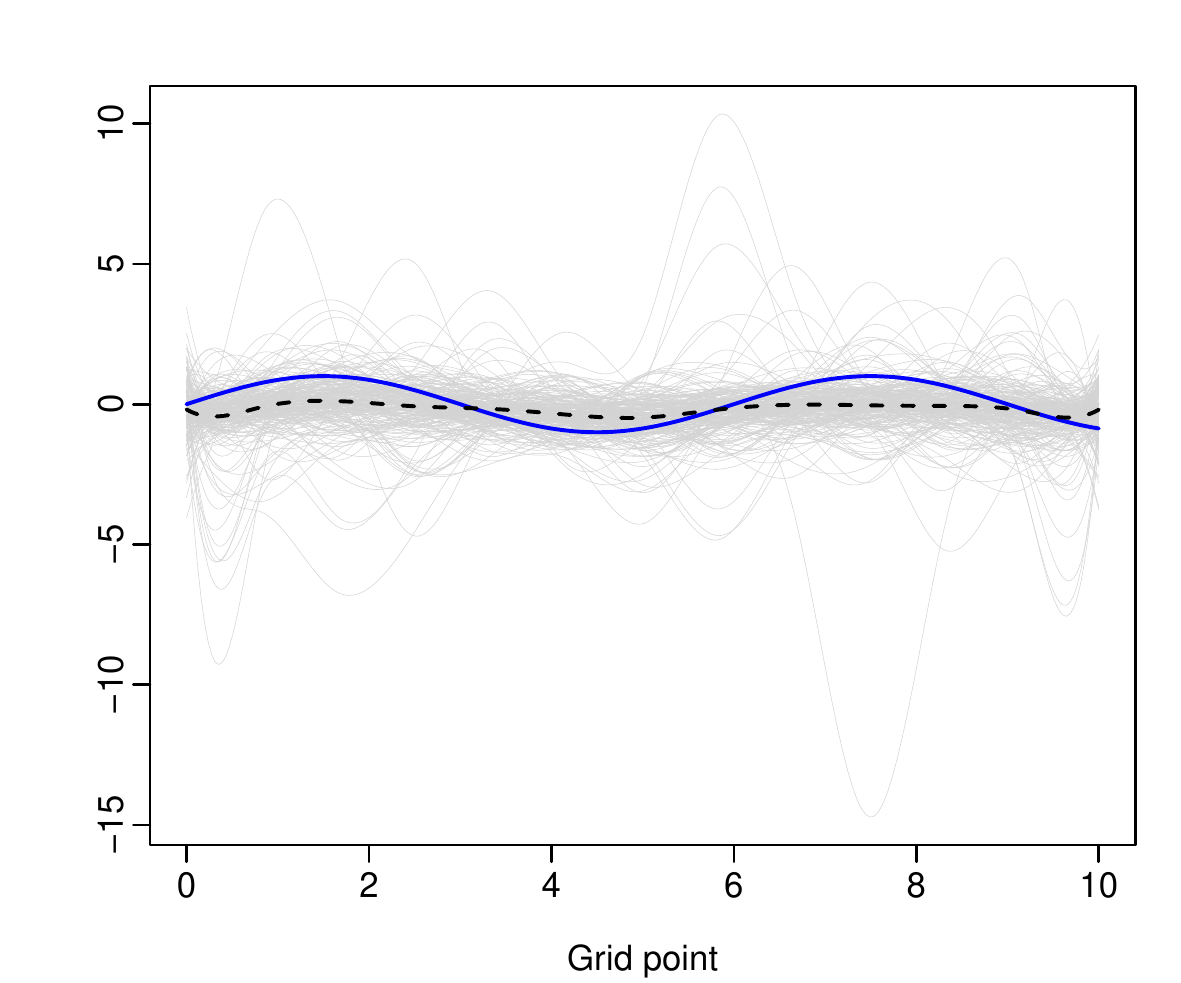}
  \includegraphics[width=5.9cm]{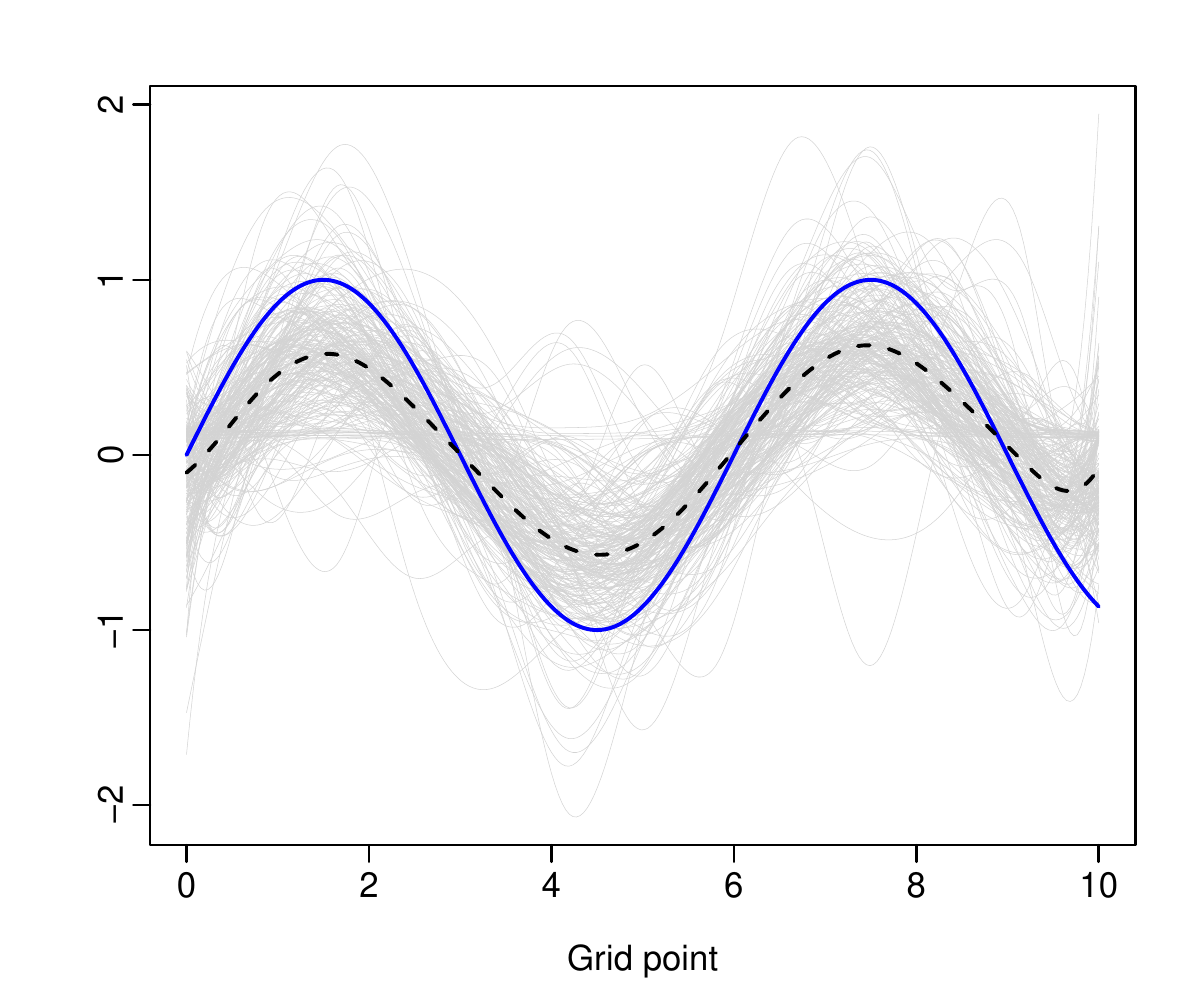}
  \caption{\small{Plots of the generated true parameter functions (blue lines), estimated parameter functions (gray lines), and mean functions of the estimated parameter functions (black lines). FPC (first column), FPLS (second column), and RFPLS (third column). The data are generated under Case 1 when no outlier is present in the data (first row), $[1\%,~5\%,~10\%]$ of the generated data are contaminated by outliers (second row, third row, and fourth row, respectively.)}}
  \label{fig:Fig_5}.
\end{figure}

Our results are presented in Figures~\ref{fig:Fig_3}-\ref{fig:Fig_5}. From Figures~\ref{fig:Fig_3} and~\ref{fig:Fig_6}, when no outlier is present in the data, the proposed method produces similar or slightly smaller CCR and AUC values compared with the traditional FPLS and FPC methods. On the other hand, when outliers are present in the data, our proposed method significantly improves CCR and AUC values than FPLS and it produces competitive or even better CCR and ACU values compared with the FPC. Especially, the difference between the classification performance of the proposed method and traditional FPLS becomes more prominent as the contamination level increases. Compared with FPC, the FPLS is generally more affected by outliers and produces smaller CCR and AUC values.

The results for the computed IMSE under Case 1 are presented in Figure~\ref{fig:Fig_4}. This figure shows that our proposed method produces better parameter estimates than the FPC and FPLS for all scenarios. Note that our proposed method is expected to produce similar IMSE values when no outlier is present in the data. However, it produces improved IMSE values even in this case. This result is because the considered data generation process may generate some outliers with small or moderate magnitudes. The proposed method downweighs such observations and produces smaller IMSE values than other methods. In addition, the plots of the generated true regression coefficient function and its estimates by the methods are presented in Figure~\ref{fig:Fig_5}. This figure also supports our findings presented in Figure~\ref{fig:Fig_4} and further confirm that our proposed method produces improved parameter estimates in all scenarios compared with the FPC and FPLS.

\subsection{Empirical data example: Strawberry puree data} 

\begin{figure}[!htb]
  \centering
  \includegraphics[width=5.9cm]{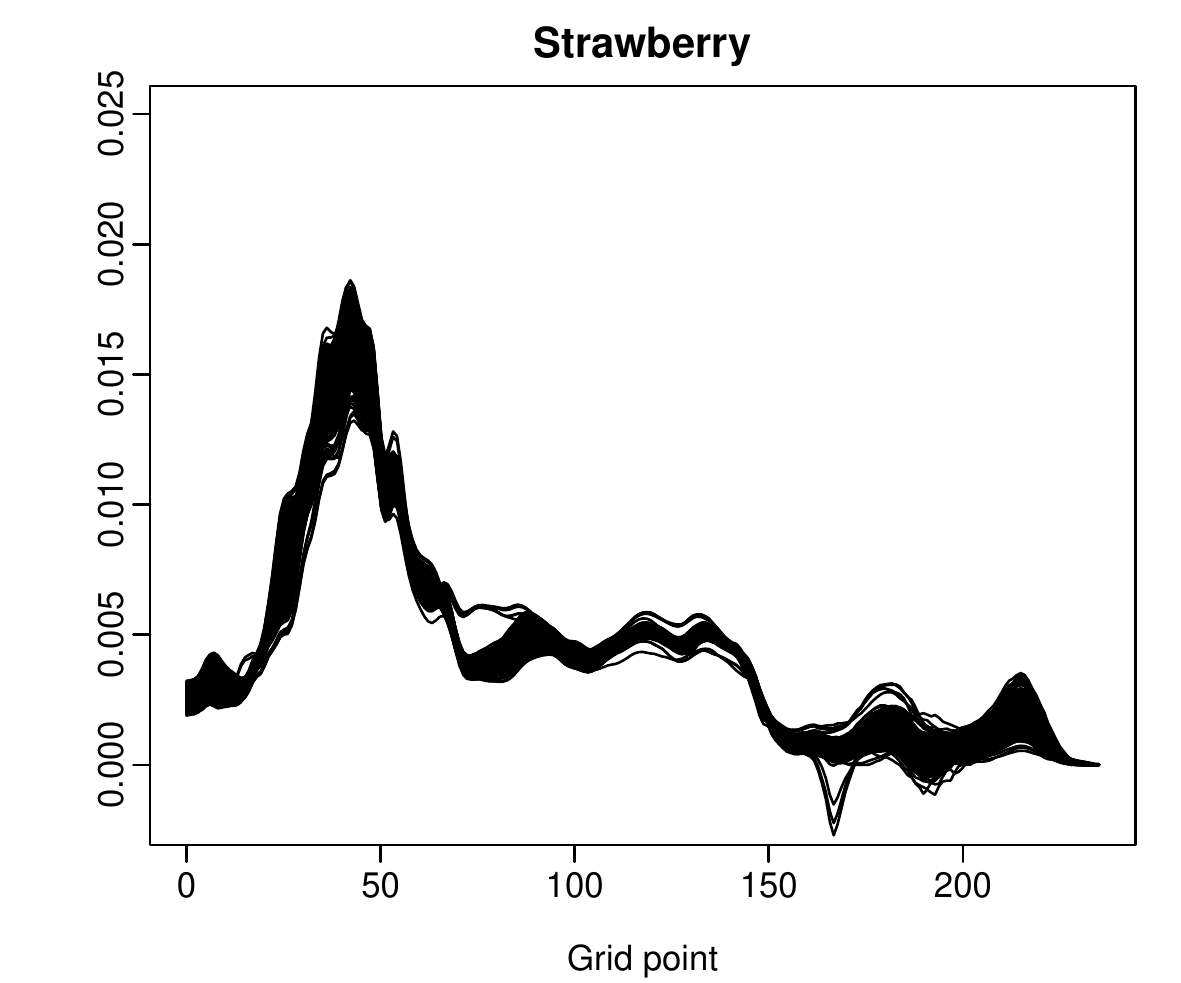}
  \includegraphics[width=5.9cm]{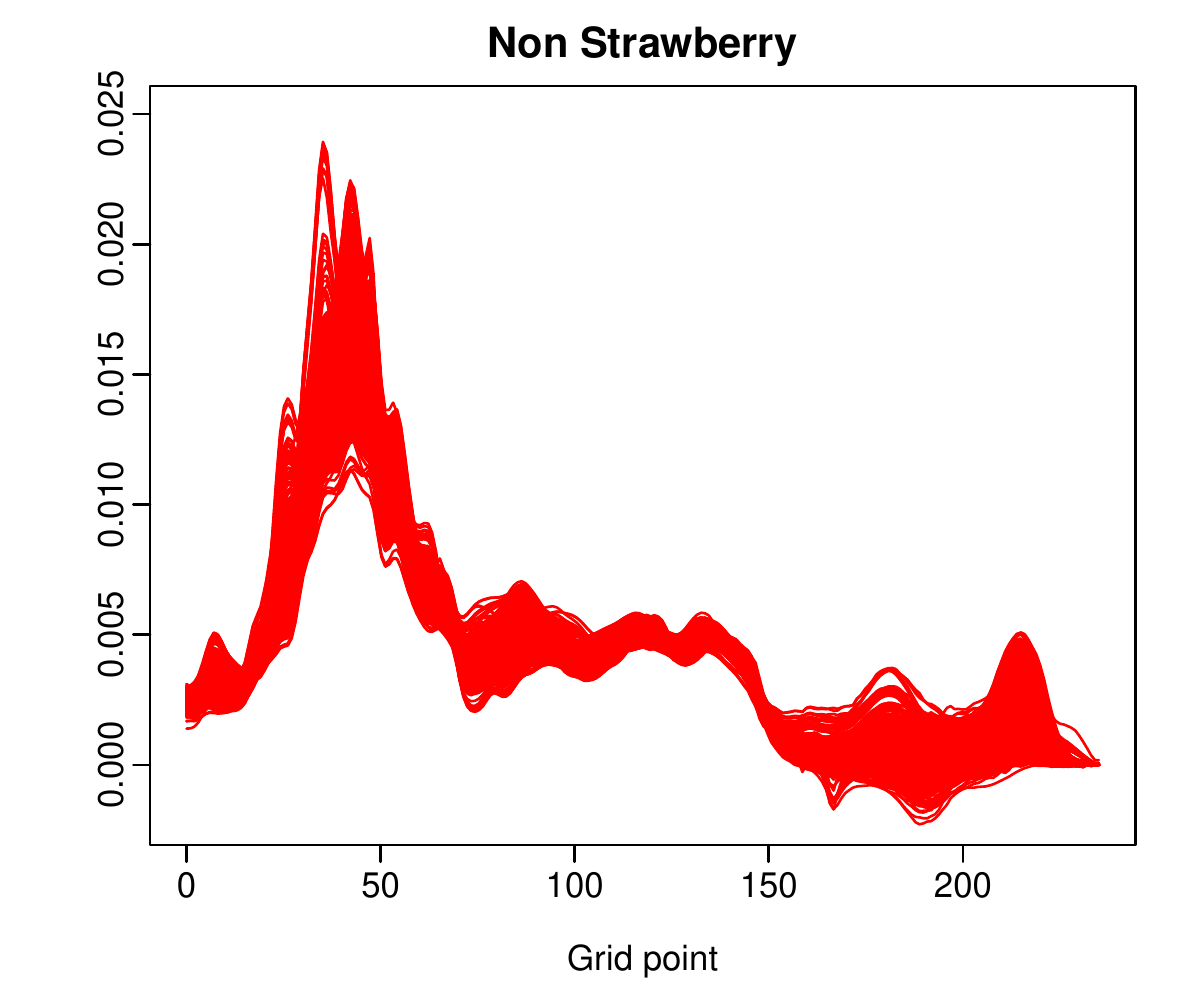}
  \includegraphics[width=5.9cm]{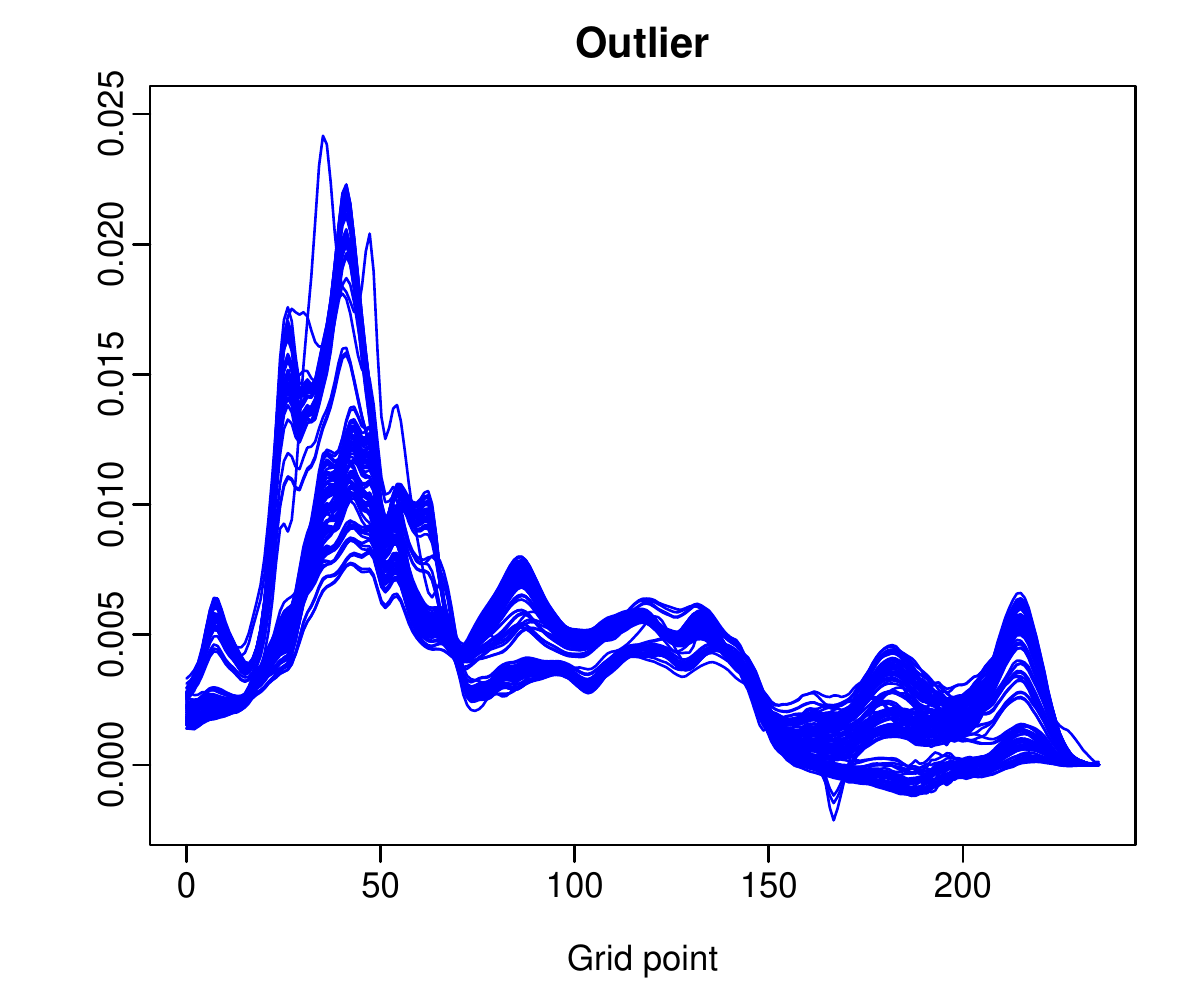}
  \caption{\small{Graphical display of the mid-infrared spectra of strawberry fruit purees (left panel), non-strawberry fruit purees (middle panel), and detected outliers (right panel)}.}
  \label{fig:Fig_7}
\end{figure}

The strawberry puree data available at \url{https://csr.quadram.ac.uk} include a total of $983$ mid-infrared spectrum collections, where $351$ of which are pure strawberries (authentic samples) and $632$ of which are non-pure strawberries (adulterated strawberries and other fruits). The fruit purees were collected using the Fourier Transform Infrared spectroscopy with attenuated total reflectivity sampling on the integrated spectral area over the range $899~\text{cm}^{-1} - 1802~\text{cm}^{-1}$ ($235$ points in total) \citep[see][for a more detailed description of the data set]{Holland1998}. We aim to classify the fruit purees (strawberry or non-strawberry) based on the mid-infrared spectra of fruit purees. For this data set, the presence of outliers is investigated via the functional boxplot method proposed by \cite{SunG} (available in the \Rlogo\ package ``\texttt{fdaoutlier}'' \citep{fdaoutlier}). According to the functional boxplot method, the functional predictor includes 122 (12.4\%) outliers (see Figure~\ref{fig:Fig_7}) with small and large magnitudes, which may lead to poor classification performance for the non-robust methods. The presence of a relatively large amount of outliers in this data set motivates us to apply our proposed method to classify the fruit purees robustly.

\begin{figure}[!htb]
  \centering
  \includegraphics[width=12cm]{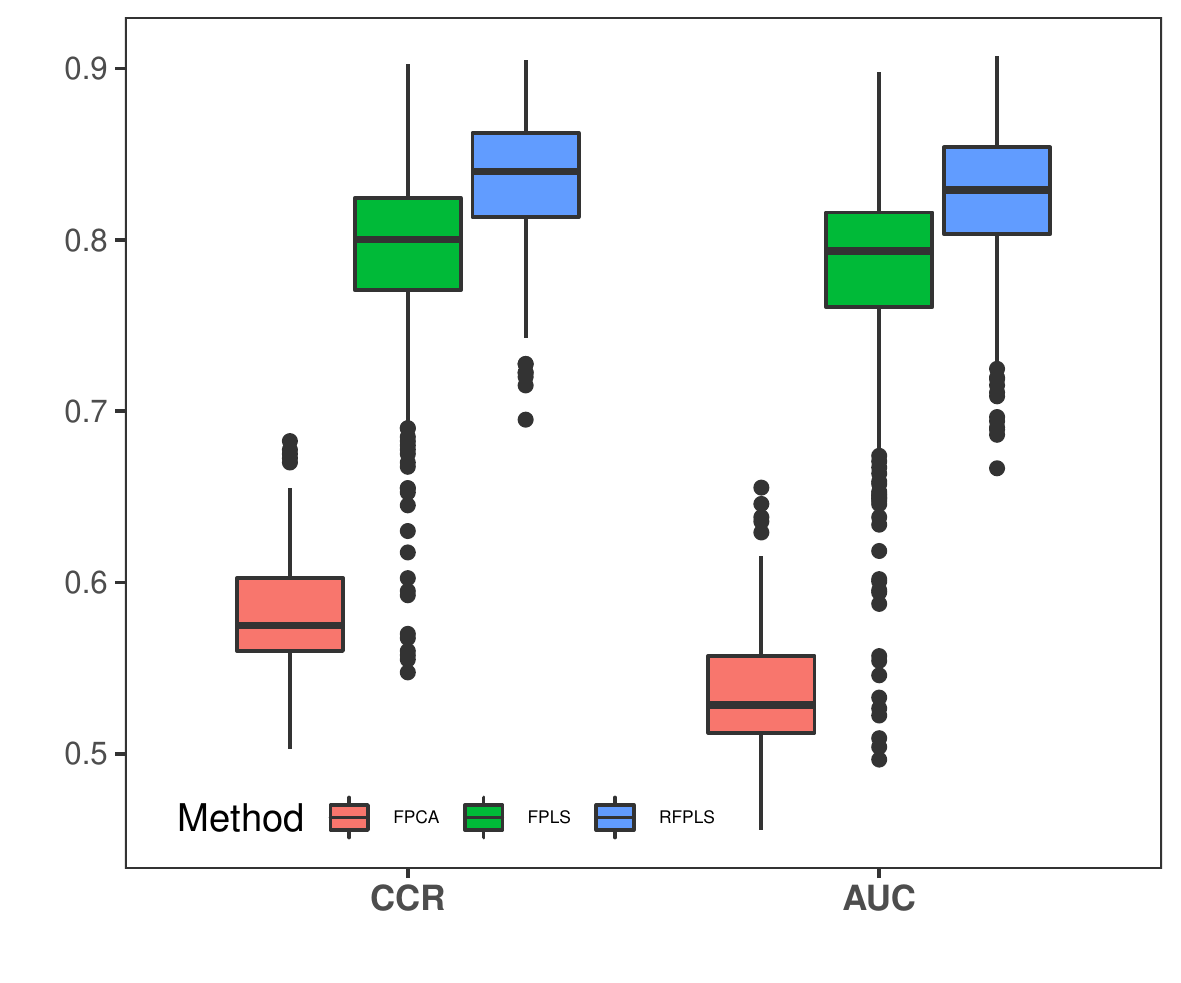}
  \caption{\small{Boxplot of the computed CCR and AUC values for the strawberry puree data}.}
  \label{fig:Fig_8}
\end{figure}

The following procedure is repeated 250 times to compare the classification performance of the methods. The entire data set is divided into a training sample with size $n_{train} = 583$ and a test sample with size $n_{test} = 400$. The models are constructed with the training sample to classify the fruit purees in the test sample based on the CCR and AUC. Note that $K = 15$ basis expansion functions are used to estimate the constructed methods. Our results are presented in Figure~\ref{fig:Fig_8}. This figure shows that the proposed method significantly outperforms the FPLS and FPC so that it produces improved CCR and AUC values compared with the FPLS and FPC. In other words, the non-robust methods (FPLS and FPC) are affected by the outliers and produce misclassification results for the binary response variable. On the other hand, our proposed method improves classification accuracy for the response variable by downweighing the effects of the outliers.

\section{Conclusion} \label{sec:conc}

Several methods have been proposed to estimate the regression coefficient function of the FLogR model. However, most of these methods are based on the LS-type estimator, and outliers can considerably hinder those estimators. In the case of outliers, the LS-based estimation methods may produce biased estimates for the regression coefficient function of the FLogR model and an increased probability of misclassification. We propose an RFPLS method to estimate the regression coefficient function of the FLogR model robustly in the presence of outliers. In the proposed method, a weighting approach is applied to the basis expansion coefficients of the functional predictor to downweigh the effects of outliers in the predictor variable. In addition, a weighted likelihood is used to compute the components of the FPLS method robustly and estimate the approximate model. The estimation and classification performance of the proposed method is evaluated by a series of Monte Carlo experiments and an empirical data analysis studying strawberry purees. The results produced from our analyses demonstrate that, compared with the existing non-robust methods, the proposed method produces improved estimates for the regression coefficient function of the FLogR model in the presence of outliers. In addition, it produces improved classification results for the binary response variable when outliers contaminate the data. Our results also demonstrate that the proposed method produces competitive performance to its traditional competitors even when no outlier is present in the data.

There are several possible directions that the proposed method in the current paper may be extended, and we outline three possibilities. First, we consider only binary outcomes in the FLogR model in the present paper. The proposed method can easily be extended to the multi-class classification problems as an alternative to \cite{Miroslav2017}, and \cite{Aguilera2019}. Second, in the current paper, we only consider functional predictors in the model. However, the classification of the binary response variable in many empirical applications (especially in health-related fields ) may be strongly dependent on scalar predictors, such as weight, height, and age. The proposed method can also be extended to the partially functional logistic regression model, where both scalar and functional predictors are used to classify the binary outcome. Lastly, we consider only one functional predictor in the model in the proposed method. The proposed method can easily be extended to multiple FLogR models when more than one functional predictor is needed in the model.

\section*{Acknowledgments}

This work was supported by The Scientific and Technological Research Council of Turkey (TUBITAK) (grant no: 120F270). The first author acknowledges the financial support from TUBITAK.

\newpage
\bibliographystyle{agsm}
\bibliography{rfq}

\end{document}